\documentclass[numberedappendix]{emulateapj}  
\usepackage[hyperfootnotes=false,naturalnames=true,letterpaper]{hyperref}
\hypersetup{pdfauthor={Dan Coe}, pdftitle={LensPerfect}}
\let\chapter=\section  

\slugcomment{Accepted for publication in ApJ}

\shorttitle{LensPerfect}
\shortauthors{Coe et al.}

\begin{document}

\title{LensPerfect: Gravitational Lens Massmap Reconstructions
Yielding Exact Reproduction of All Multiple Images}

\author{D.~Coe\altaffilmark{1,2,3}}
\author{E.~Fuselier\altaffilmark{4}}
\author{N.~Ben{\'i}tez\altaffilmark{5,3}}
\author{T.~Broadhurst\altaffilmark{6}}
\author{B.~Frye\altaffilmark{7}}  
\author{H.~Ford\altaffilmark{2}}
\email{coe@caltech.edu}

\altaffiltext{1}{
Jet Propulsion Laboratory, 
California Institute of Technology, 
MS 169-327, Pasadena, CA 91109}
\altaffiltext{2}{
Johns Hopkins University, 
Department of Physics \& Astronomy, 
3400 N. Charles St., Baltimore, MD 21218, USA}
\altaffiltext{3}{
Instituto de Astrof\'isica de Andaluc\'ia (CSIC), 
Camino Bajo de Hu\'etor 50, Granada 18008, Spain}
\altaffiltext{4}{
United States Military Academy,
Department of Mathematical Sciences,
West Point, NY 10996}
\altaffiltext{5}{
Instituto de Matem\'aticas y F\'isica Fundamental (CSIC),
C/ Serrano 113-bis, Madrid 28006, Spain}
\altaffiltext{6}{
Tel Aviv University,
School of Physics and Astronomy,
Tel Aviv 69978, Israel}
\altaffiltext{7}{
Dublin City University,
Department of Physical Sciences,
Dublin 9, Ireland}

\begin{abstract}
We present a new approach to gravitational lens massmap reconstruction.
Our massmap solutions perfectly reproduce
the positions, fluxes, and shears of all multiple images.
And each massmap accurately recovers the underlying mass distribution
to a resolution limited by the number of multiple images detected.
We demonstrate our technique given a mock galaxy cluster similar to Abell 1689
which gravitationally lenses 19 mock background galaxies 
to produce 93 multiple images.
We also explore cases in which as few as four multiple images are observed.
Massmap solutions are never unique, 
and our method makes it possible to explore 
an extremely flexible range
of physical (and unphysical) solutions,
all of which perfectly reproduce the data given.
Each reconfiguration of the source galaxies produces a new massmap solution.
An optimization routine is provided to find those source positions 
(and redshifts, within uncertainties)
which produce the ``most physical'' massmap solution,
according to a new figure of merit developed here.
Our method imposes no assumptions about 
the slope of the radial profile nor mass following light.
But unlike ``non-parametric'' grid-based methods,
the number of free parameters we solve for 
is only as many as the number of observable constraints
(or slightly greater if fluxes are constrained).
For each set of source positions and redshifts,
massmap solutions are obtained ``instantly'' via direct matrix inversion
by smoothly interpolating the deflection field
using a recently developed mathematical technique.
Our LensPerfect software is straightforward and easy to use
and is made publicly available via our website.

\end{abstract}


\pagenumbering{arabic}  
\section{Introduction}
\label{intro}

Simulations of structure formation in a $\Lambda$CDM universe 
have provided concrete predictions for 
the form of Dark Matter halos over a wide range of scales
\citep[e.g.,][]{Merritt06, Bett07, Diemand07}.
These mass distributions,
quantified in terms of radial profile, ellipticity, and level of substructure,
are among the key predictions of $\Lambda$CDM theory.
Uncertainties do persist, 
especially with regard to baryons, absent from most Dark Matter simulations.
Baryons are found to alter the inner profiles and ellipticities of halos,
especially on galaxy scales
\citep[e.g.,][]{Kazantzidis04, Gustafsson06, Dutton07}.

Testing these predictions in detail observationally has proven challenging.
Gravitational lensing provides us with our most direct tool
for mapping the distributions of mass (predominantly Dark Matter)
within and surrounding galaxies and galaxy clusters.
But massmaps recovered by this method 
are of much lower resolution than simulated Dark Matter halos.
Improvements in imaging quality both from the ground and space
now allow for a more definitive measurement of lensing effects,
motivating new techniques to take full advantage of this advance.

Deep multi-wavelength high-resolution images of five galaxy clusters
(Abell 1689, Abell 2218, Abell 1703, MS1358, and CL0024)
have been obtained with ACS, the Advanced Camera for Surveys \citep{ACS}
on-board the Hubble Space Telescope.
The deepest of these, taken of A1689, 
has revealed an unprecedented number of multiple images,
with 106 images of 30 source galaxies identified in the original analysis
by \cite{Broadhurst05}.
This should allow one to reconstruct a relatively high resolution massmap
of the cluster's Dark Matter halo.

\begin{figure*}
\plottwo{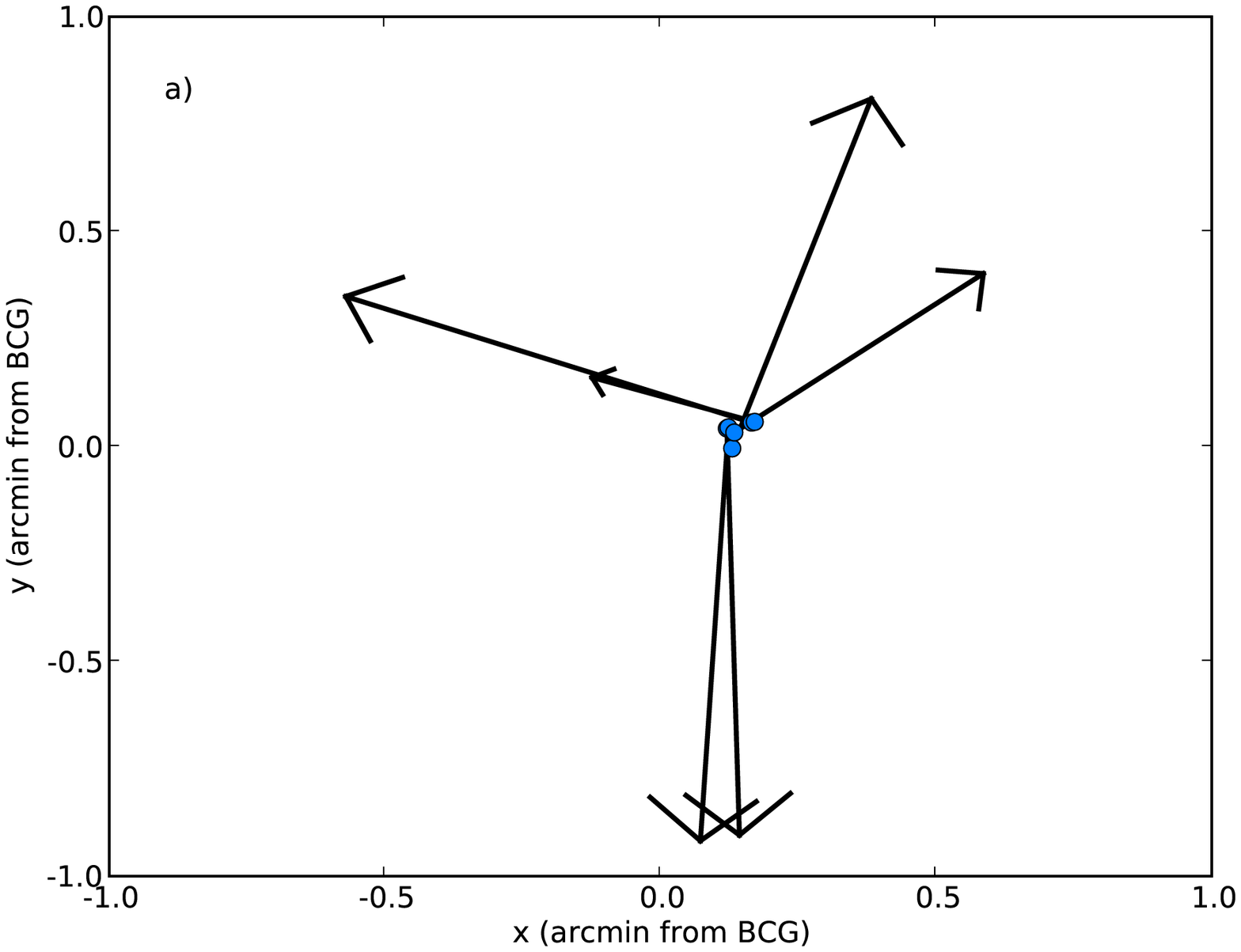}{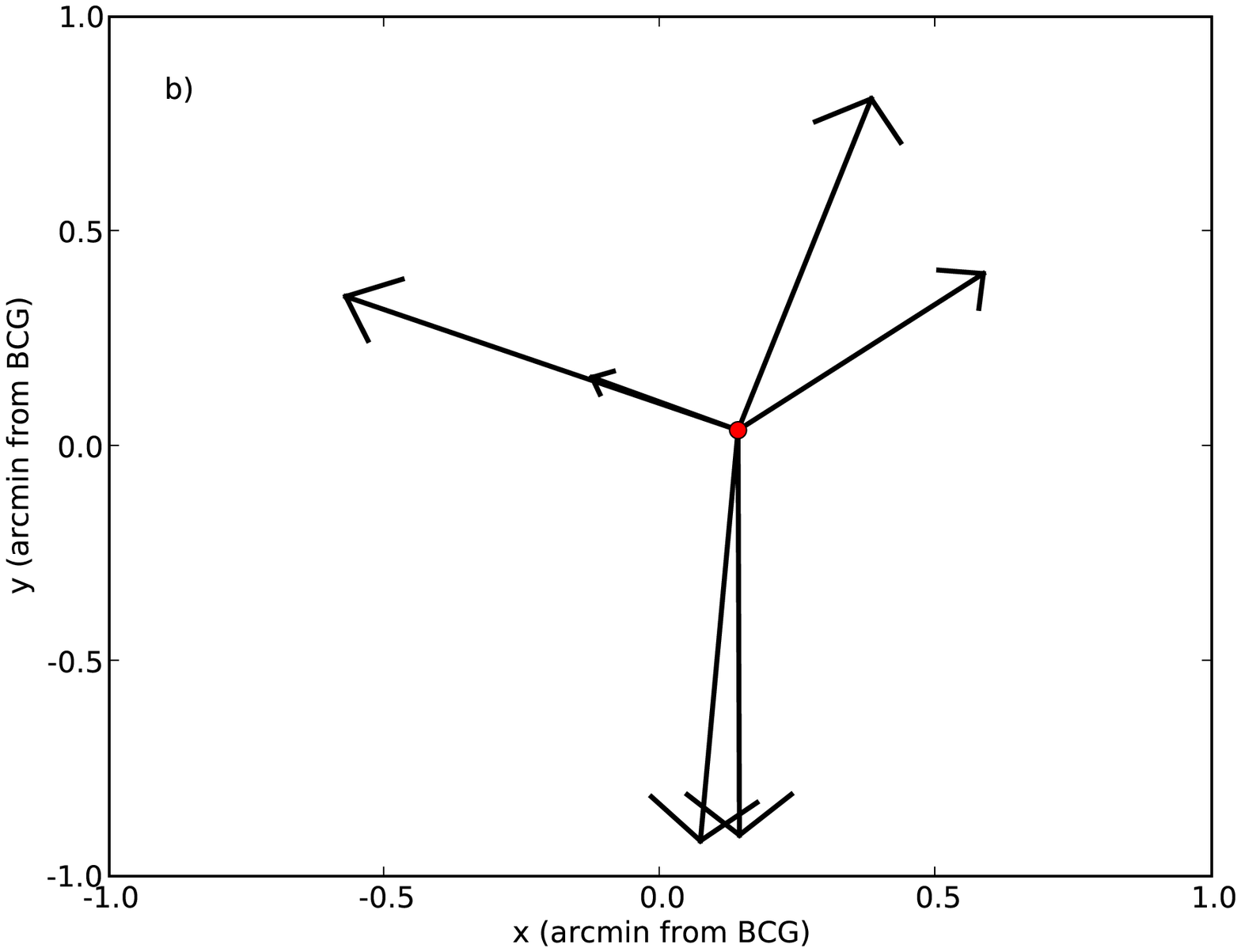}
\caption[Imperfect vs.~perfect reconstruction]{
\label{delens1}
Left: Multiple images of object \#1 from \citep{Broadhurst05} 
delensed back to the source plane
using their best fit deflection field.
The scatter in the source plane ($\sim 2\arcsec$)
is typical of all model-based massmap reconstruction methods used to date.
Right: The same multiple image positions
all delensed back to a single source position.
These deflections can be fit exactly by a LensPerfect solution.
}
\end{figure*}

\cite{Broadhurst05} used a novel massmap parameterization to
reproduce the 106 observed multiple images well but not exactly
(Fig.~\ref{delens1}a),
with positional offsets of 3\farcs2 RMS on average in the image plane.
Note that these are very significant offsets,
as image positions can typically be determined to $0\farcs05$
(the resolution of ACS pixels) or better.
Subsequent analyses using similar methods
(with different parameterizations for the massmap
and revised multiple image lists)
improved only marginally on these positional offsets to 2\farcs5 RMS at best
\citep{Zekser06, Halkola06, Halkola07, Limousin07}.
It is possible that cosmic variation of mass density
along the multiple sight lines 
is responsible for some of this scatter.
But it is more likely that line of sight variations
are an additional source of error (present in any lensing analysis)
on top of the models' inability to properly reproduce all the image positions.

The aforementioned analyses all employ extensions of techniques
developed for use with far fewer multiple images.
A simple parametric model consisting of 
an elliptical mass profile plus external shear
can accurately reproduce the positions of four multiple images 
\citep[e.g.,][]{Keeton97}.
But such models struggle when confronted with additional multiple images.
Multi-component (``halo'' + ``galaxy'') models
are required to obtain decent fits to galaxy cluster lensing
\citep[e.g.,][]{Kneib96, Broadhurst00}.
The A1689 multi-component models discussed above
don't contain enough free parameters to perfectly fit the data.
Yet they already contain so many parameters
that navigating the parameter space proves challenging.
It may be impossible to find the best solution with so many free parameters
given our current computational capabilities.
Advances in computing power may someday allow parametric models
the freedom necessary to produce perfect fits.
Allowing galaxy subhalos to drift in position and vary individually in mass
could dramatically improve the fits
while breaking free from the assumption 
that Dark Matter subhalos strictly follow the light distribution.

The degree to which Dark Matter follows light (and/or gas)
is an important outstanding question.
The exciting discoveries in this area
are currently being made by {\em weak} gravitational lensing analyses
which make no assumptions about the underlying mass distributions.
The ``Bullet Cluster'' finding \citep{Clowe06}
that gas has been stripped from two colliding Dark Matter halos
would not have been possible
had the authors assumed mass follows light from the outset.
A similar cluster collision observed along the line of sight
appears to have left a Dark Matter ring around CL0024.
This ring was also detected by weak lensing analysis \citep{Jee07}.

For many years now, 
we have obtained assumption-free massmap reconstructions
from direct inversion of weak gravitational lensing data
\citep{Tyson90, KaiserSquires93}.
Here we present the first method to do the same given
{\em strong} gravitational lensing data (multiple images).
Our method is not entirely assumption free, 
as a few basic assumptions about the distribution of mass
can be helpful
in selecting the most ``physical'' among the possible massmap solutions
(\S\ \ref{sourcepos}).

Model-free massmap reconstructions have previously been obtained for Abell 1689
using strong lensing data (though not via direct inversion).
As these massmaps are more flexible than model-based solutions,
they should produce better fits to the data.
But this promise has yet to be fully realized.
The SLAP method \citep{Diego05a} computes massmaps on an adaptive mesh grid,
and fits the data to a desired level of scatter.
But when this level is set too low,
the solutions become ``biased'' with ``a lot of substructure''.
Their best solution for A1689
leaves scatters of $\sim 3\arcsec$ in the {\em source} plane
(\citealt{Diego05}; Diego 2007, private communication).
Meanwhile, \cite{Saha06} use a method called PixeLens
to obtain pixel-based (fixed grid) massmaps which perfectly reproduce 
all multiple image positions.
But computational constraints prevent them from fitting more than 
30 images at a time.
The massmap we obtain for A1689 (to be presented in an upcoming paper)
perfectly reproduces the positions of 135 multiple images 
(as in Fig.~\ref{delens1}b)
and thus has about twice the resolution as the PixeLens massmap 
in each spatial direction
(as dictated by the density of multiple images).

Massmaps may be further improved by incorporating constraints
beyond simple image positions.
Images which are resolved and extended 
should be properly reproduced by the mass model
\citep{WarrenDye03, Suyu06, Koopmans06}.

Even unresolved images yield the information encoded in their fluxes.
A simple mass model which accurately reproduces ``quad'' image positions
may or may not accurately reproduce the image flux ratios.
Discrepant cases are known as ``flux anomalies''.
If other causes (microlensing, time delays, and dust extinction)
can be ruled out or accounted for,
then small substructure ($\sim 10^6 M_\odot$ subhalos)
is generally invoked as the most likely explanation for an observed anomaly.

But this explanation has perhaps been invoked too often.
The amount of substructure observed in simulations
may not be sufficient to produce
flux anomalies as often as observed
\citep{Metcalf04, Amara06, MaccioMiranda06, Diemand07}.
One way to resolve this possible discrepancy
is to obtain macro mass model solutions which reproduce the observed fluxes
without resorting to smaller substructure.

To address this issue,
\cite{EvansWitt03} developed a direct inversion massmap reconstruction method
capable of perfectly fitting 
the observed positions and fluxes of four multiple images.
While providing reasonable solutions for the lensed systems
Q2237+0305 and PG1115+080, their solution for B1422+231 was clearly unphysical.
While it is unclear how exactly to quantify a massmap's physicality,
the authors characterized their B1422+213 model
as simply too ``wiggly'' to be plausible.
Developing the method a bit further, \cite{CongdonKeeton05}
produced a more reasonable solution for B1422+231,
but were less successful with B2045+265 and B1933+503.
For the unsuccessful cases, the authors argue that small scale substructure
is the most likely explanation for the observed flux anomalies.
But perhaps their models
were simply not flexible enough to obtain physical solutions for these systems.
We will revisit this question in an upcoming paper.

Our new LensPerfect method uses direct inversion
to obtain assumption-free massmap solutions
which perfectly reproduce all multiple image positions.
Multiple image knots and fluxes may also be perfectly constrained.
LensPerfect is made possible by a recent advance in the field of mathematics.
The essential tool is a method that produces a curl-free interpolation
of vectors given at scattered data points.
A set of observed multiple image positions and redshifts
along with assumed source positions
defines a deflection field at the image positions.
Once we interpolate this deflection field across the entire field,
we take the gradient, multiply by 2, 
and ``instantly'' obtain our perfect massmap solution.

The interpolation of data given at scattered points is a complex problem
without a unique solution.
One method of attacking such problems involves the use of
radial basis functions (RBFs).
RBFs have been used to interpolate scattered data 
since the early 1970's \citep{Hardy71}
and are used in many applications today.
By the 1980's, RBFs had been applied to interpolate vector fields.
And in 1994, a method was developed capable of yielding
divergence-free interpolations of scattered vectors \citep{Narcowich94}.
Finally, the curl-free analog of this method was developed last year
\citep{Fuselier06, Fuselier07}.
Here we apply this new method to gravitational lensing analysis.

We describe our method
along with the necessary gravitational lensing equations in \S\ref{sec:method}.
In \S\ \ref{sec:applications}, 
we demonstrate applications of the method, including
the recovery of a known massmap given 93 multiple images (\S\ \ref{A1689model}).
We also explore the solutions obtained 
when only a handful of multiple images are available (\S\ \ref{Einstein}).
And we demonstrate the gains made by 
constraining extra image knots in \S\ \ref{knots}.
The method for adding flux and/or shear constraints 
is given in \S\ \ref{fluxes}.
In \S\ \ref{discussion}
we discuss the relative merits of model-based and model-free methods,
along with the potential of a hybrid method 
among other techniques that may become possible with LensPerfect.
Finally, we provide a summary in \S\ \ref{summary}.
The LensPerfect software and more information are available at our 
website.\footnote{http://www.its.caltech.edu/\%7Ecoe/LensPerfect/}

\section{Method}
\label{sec:method}

\begin{figure}
\epsscale{0.7}
\plotone{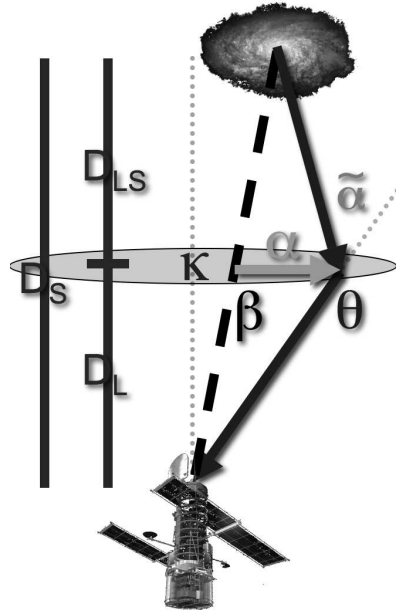}
\epsscale{1.0}
\caption[Deflection of light due to a gravitational lens]{
\label{deflection}
Light ray deflection by a gravitational lens 
of mass $\kappa$. In the absence of $\kappa$,
the galaxy would appear at its true, or ``source'', position $\vec\beta$.
The intervening mass deflects its light 
by an amount $\vec\alpha$ to position $\vec\theta$.
The deflection angle $\alpha$ on our sky
is related to the actual bend angle $\tilde \alpha$ of the light ray
via $\alpha D_S = \tilde \alpha D_{LS}$.
$D_S$, $D_{LS}$, and $D_L$ are measured as angular diameter distances.
}
\end{figure}

\subsection{Image Deflection}

Image deflection by a gravitational lens is governed by 
a few simple equations \citep[e.g.,][]{Wambsganss98}.
The relativistic bending of light 
due to a mass $M$ at a distance $R$ away
is twice that expected from Newtonian physics: 
$\tilde \alpha = 4 G M / c^2 R$
given Newton's gravitational constant $G$,
and the speed of light $c$.
In a gravitational lens, it is generally safe to assume
that all of the deflection occurs in the plane of the lens.
(This is known as the thin lens approximation.)
Given the projected mass surface density distribution
$\kappa(\vec \theta)$ of the lens,
we can derive the deflection of light 
$\vec \alpha (\vec \theta) = \vec \theta - \vec \beta$
from its true position on the sky $\vec \beta$ 
to that at which we observe it $\vec \theta$ (Fig.~\ref{deflection}):
\\
\begin{equation}
  \label{deflkappa}
  \vec \alpha (\vec \theta) = \frac{1}{\pi} \int d^2 \vec \theta \arcmin
  \kappa(\vec \theta \arcmin)
  \frac{\vec \theta - \vec \theta \arcmin}
  {\left \vert \vec \theta - \vec \theta \arcmin \right \vert ^2},
\end{equation}
\\
with the corresponding less-intimidating inverse relation:
\\
\begin{equation}
\label{div}
\nabla \cdot \vec \alpha = 2 \kappa.
\end{equation}
The surface density $\kappa = \Sigma / \Sigma_{crit}$ 
is defined in units of the critical density at the epoch of the lens.
The critical density is that generally required 
for multiple images to be produced.
It is a function of source redshift as given by:
\\
\begin{equation}
\label{E_c}
\Sigma_{crit} = \frac{c^2}{4 \pi G} \frac{D_S}{D_L D_{LS}},
\end{equation}
\\
involving a ratio of the angular-diameter distances from
observer to source $D_S = D_A(0,z_S)$,
observer to lens $D_L = D_A(0,z_L)$,
and lens to source $D_{LS} = D_A(z_L,z_S)$.
For a flat universe ($\Omega = \Omega_m + \Omega_\Lambda = 1$), 
angular-diameter distances are calculated as follows 
\citep[filled beam approximation; see also \citealt{Hoggcosmo}]{Fukugita92}:
\\
\begin{equation}
  D_A(z_1,z_2) = \frac{c}{1 + z_2} \int_{z_1}^{z_2} \frac{dz^\prime}{H(z^\prime)},
\end{equation}
\\
where the Hubble parameter varies with redshift as:
\\
\begin{equation}
  H(z) = H_o \sqrt{\Omega_m (1 + z)^3 + \Omega_\Lambda}.
\end{equation}
\indent Thus the critical density $\Sigma_{crit}$
is a function of the source redshift.
This follows because the deflection angle $\vec \alpha$ 
is a function of source redshift.
As source redshift decreases, 
the light bend angle $\tilde \alpha$ remains constant, which
(imagine moving the galaxy in Fig.~\ref{deflection} 
inward along the top blue arrow)
requires the image deflection to decrease by the distance ratio:
\\
\begin{equation}
\label{deflscale}
\vec \alpha = \left( \frac{D_{LS}}{D_S}  \right) \vec \alpha_\infty,
\end{equation}
\\
where $\vec \alpha_\infty$ is the deflection for a source at infinite redshift.

Thus the problem of massmap reconstruction
can be reduced to determining the deflection field
with all deflections scaled to a common redshift (e.g., $\vec\alpha_\infty$),
at which point we simply take the divergence 
and divide by 2 to obtain the massmap (Eq.~\ref{div}).
The deflection field
$\vec \alpha(\vec \theta) = \vec \theta - \vec \beta$
may be measured at the multiple image positions $\vec \theta$
once source positions $\vec \beta$ are determined (see \S\ \ref{sourcepos}).
However, in order to take its divergence,
the deflection field must be solved for
as a continuous function of position
(or at least defined on a regular grid).
Our interpolated deflection field must also be curl-free:
\\
\begin{equation}
\label{curlfree}
\nabla \times \vec \alpha = 0.
\end{equation}
This follows from Equation \ref{deflkappa}
in the same way that we find
an electric field due to a static distribution of charge is also curl-free.
One way to demonstrate this is 
to use the substitution $\nabla \ln \theta = \vec \theta / \theta^2$
to define the lensing potential:
\\
\begin{equation}
  \label{potentialkappa}
  \psi (\vec \theta) = \frac{1}{\pi} \int d^2 \vec \theta \arcmin
  \kappa(\vec \theta \arcmin)
  \ln \left \vert \vec \theta - \vec \theta \arcmin \right \vert,
\end{equation}
\\
such that
\\
\begin{equation}
  \vec \alpha = \nabla \psi.
\end{equation}
\\
The fact that the deflection field $\vec \alpha (\vec \theta)$
may be written as the gradient of a scalar field $\psi (\vec \theta)$
guarantees that it has no curl.

\subsection{Curl-Free Vector Interpolation}
\label{}

\citet{Fuselier06, Fuselier07} 
has shown how to obtain a curl-free interpolation 
of a vector field given on scattered data points.
The method is general enough to be applied to an arbitrary number of 
dimensions, but will be discussed here for the 2-D case.
The interpolated vector field is constructed using Radial Basis Functions, 
or RBFs.
An RBF is a positive definite function with radial symmetry $\phi(R/R_o)$.
The fact that it is positive definite\footnote{An $m \times m$ matrix-valued
function $\phi$ is {\em positive definite} on $\mathbb{R}^n$ 
if given any finite, distinct set of points 
$X := \{ x_1, \dots, x_N \} \subset \mathbb{R}^n$ we have
$\displaystyle \sum_{j,k} \alpha^T_j \phi(x_j - x_k) \alpha_k \geq 0$
for all $\alpha_1, \dots, \alpha_N$ in $\mathbb{R}^m$.}
guarantees that our interpolation matrix (below) will have a solution
\citep{Wendland05, Fuselier06}.
The scale length $R_o$ is input by the user,
and as we show in \S \ref{Einstein},
its freedom is closely related to the classic ``mass-sheet'' degeneracy.

In our case we must choose smooth RBFs that are at least 
three times continuously differentiable, or ``$C^3$'',
this to ensure our final massmap has no discontinuities.
Then a set of two curl-free basis vectors 
are constructed at each data point (image position)
by simply taking derivatives of the basis function:
\\
\begin{eqnarray}  \label{basis}
\vec V_a & = & \phi_{,xx} \hat x + \phi_{,xy} \hat y,\\
\vec V_b & = & \phi_{,yx} \hat x + \phi_{,yy} \hat y,
\end{eqnarray}
\\
where $\phi_{,xx}$ is the 2nd derivative of $\phi(R/R_o)$ with respect to $x$,
etc.
The vector at that data point is given as a sum over all data points
(with coefficients to be solved for):
\\
\begin{equation}
\label{coeffs}
\vec \alpha = \sum_i^N  \left(  c_a \vec V_{a_i} + c_b \vec V_{b_i}   \right).
\end{equation}
Rewriting this in matrix form we have:
\\
\begin{equation}
\left[ \alpha \right] =
\left[ \phi_{,(xy)} \right]
\left[ c \right].
\end{equation}
For example, in the case of 2 data points, $\vec \alpha (\vec \theta_1)$ and $\vec \alpha (\vec \theta_2)$:
\\
\begin{equation}
\label{deflmatrix}
\left[
\begin{array}{c}
\alpha_{x_1}\\
\alpha_{y_1}\\
\alpha_{x_2}\\
\alpha_{y_2}
\end{array}
\right]
=
\left[
\begin{array}{cc}
{\phi_{,xx_{11}}} ~~ {\phi_{,yx_{11}}} ~~ {\phi_{,xx_{12}}} ~~ {\phi_{,yx_{12}}}\\
{\phi_{,xy_{11}}} ~~ {\phi_{,yy_{11}}} ~~ {\phi_{,xy_{12}}} ~~ {\phi_{,yy_{12}}}\\
{\phi_{,xx_{21}}} ~~ {\phi_{,yx_{21}}} ~~ {\phi_{,xx_{22}}} ~~ {\phi_{,yx_{22}}}\\
{\phi_{,xy_{21}}} ~~ {\phi_{,yy_{21}}} ~~ {\phi_{,xy_{22}}} ~~ {\phi_{,yy_{22}}}
\end{array}
\right]
\left[
\begin{array}{c}
c_{a_1}\\
c_{b_1}\\
c_{a_2}\\
c_{b_2}
\end{array}
\right],
\end{equation}
\\
where, e.g.:
\\
\begin{equation}
\phi_{,yx_{12}} = \left[ \frac{\partial^2 \phi}{\partial y \partial x} \right]_{R=R_{12}},
\end{equation}
\\
with the derivative being evaluated at
$R = R_{12} = \left\vert \vec \theta_1 - \vec \theta_2 \right\vert$.
We may now simply solve for the coefficients via the matrix inverse:
\\
\begin{equation}
\left[ c \right] =
\left[ \phi_{,(xy)} \right] ^ {-1}
\left[ \alpha \right].
\end{equation}
We may guarantee that this matrix inversion yields a solution
by selecting a positive definite function for $\phi$.
Wendland functions are especially suitable choices,
and here we use this ``$C^6$'', 
or 6 times continuously differentiable,
Wendland function known as $W_{7,3}(\xi)$:
\\
\begin{equation}
\phi(\xi) = \left\{
\begin{array}{lc}
(1 - |\xi|)^8 (32 |\xi|^3 + 25|\xi|^2 + 8|\xi| + 1),& |\xi| \leq 1\\
0, & |\xi| > 1
\end{array}\right.
\label{WendlandC6}
\end{equation}
\\
where $\xi = R / R_o$.
The function is very similar to 
a Gaussian with a peak of 1 and
FWHM $\sim 0.50 R_o$ ($\sigma \sim 0.21 R_o$).\footnote{In
principle, a Gaussian could be used in place of $W_{7,3}$.
However it is well known (to mathematicians) that the interpolation matrix
is ill-conditioned when a Gaussian is used.  
Further, it is easier for a computer to evaluate a polynomial than a Gaussian.}

\begin{figure}
\plottwo{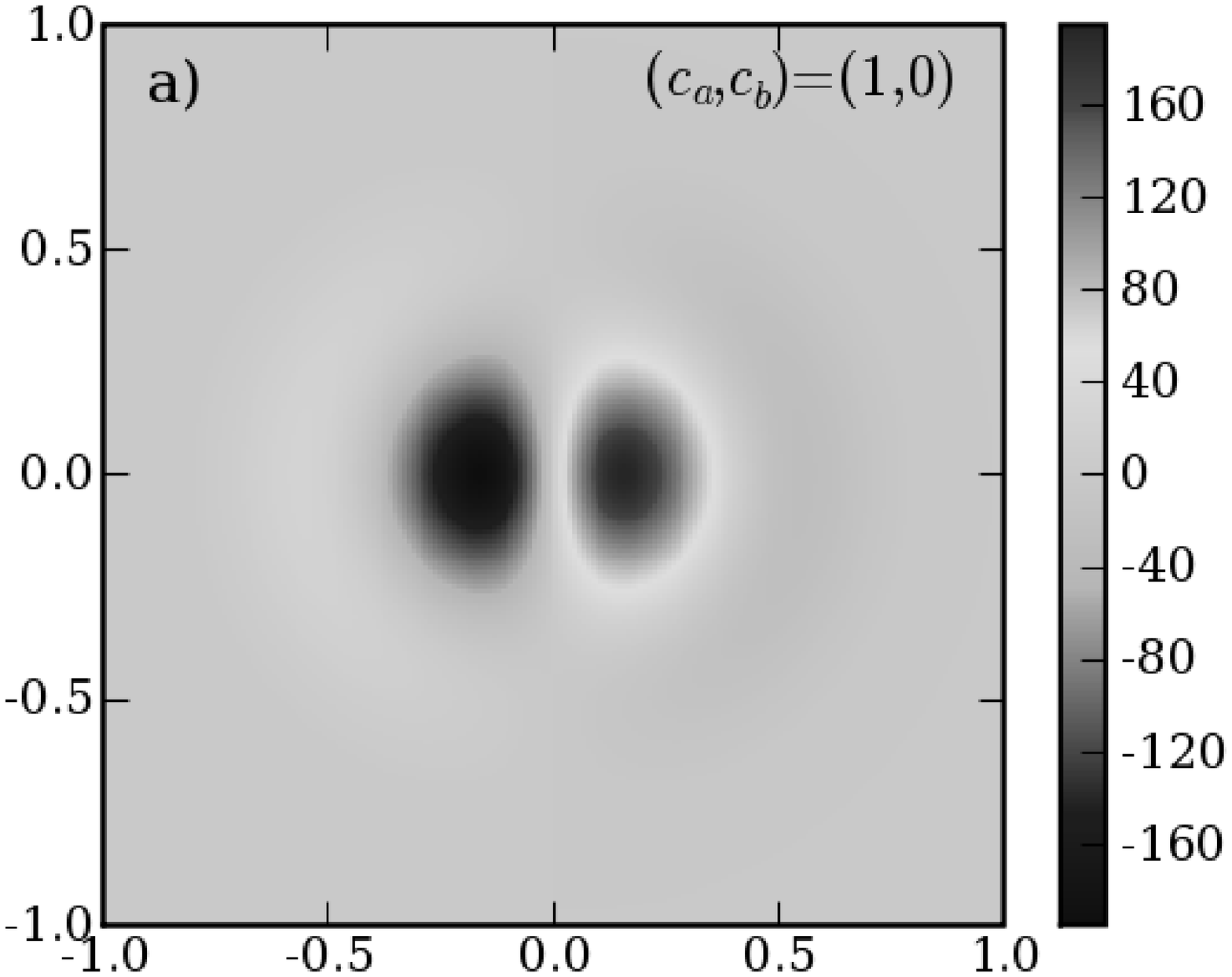}{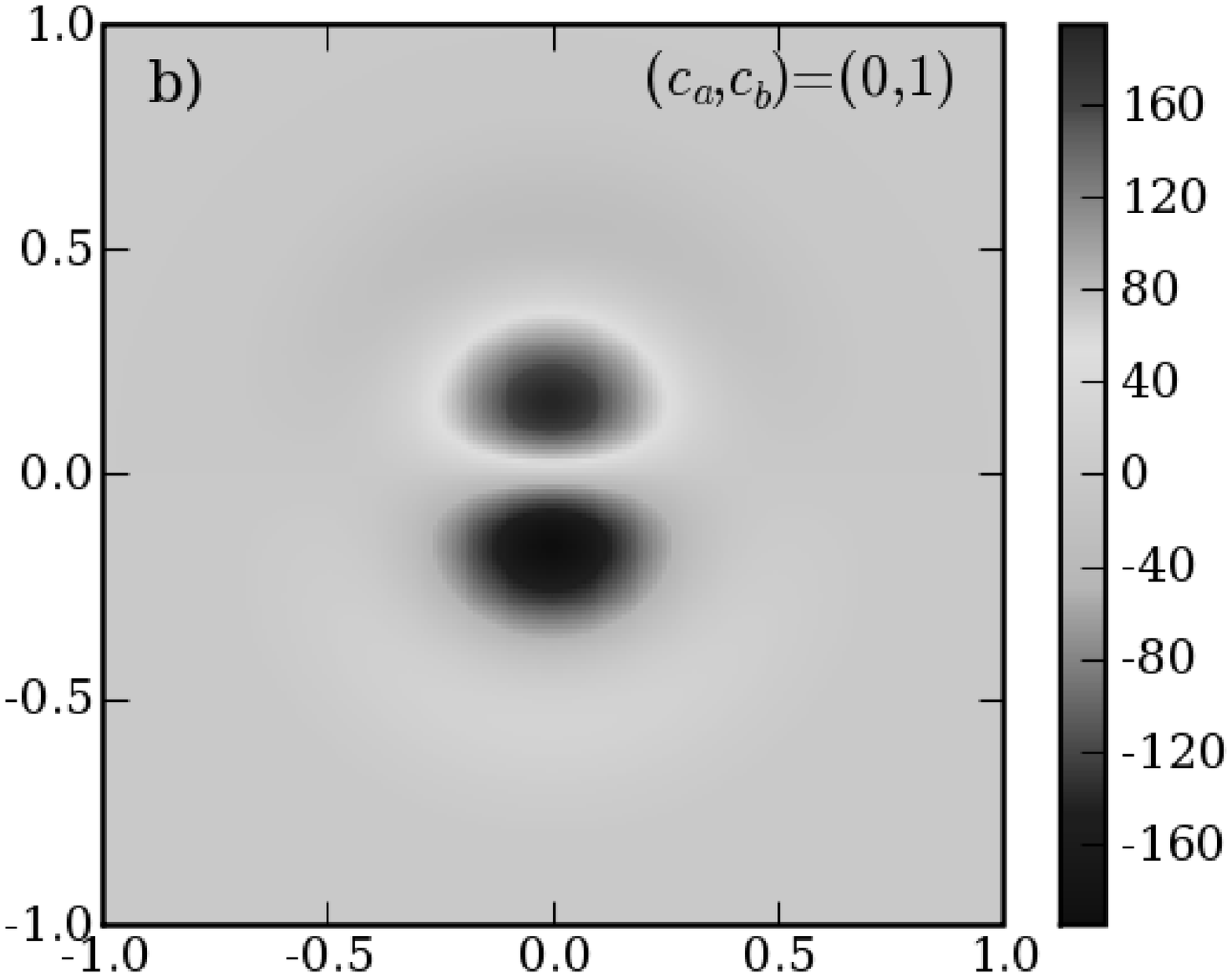}
\plottwo{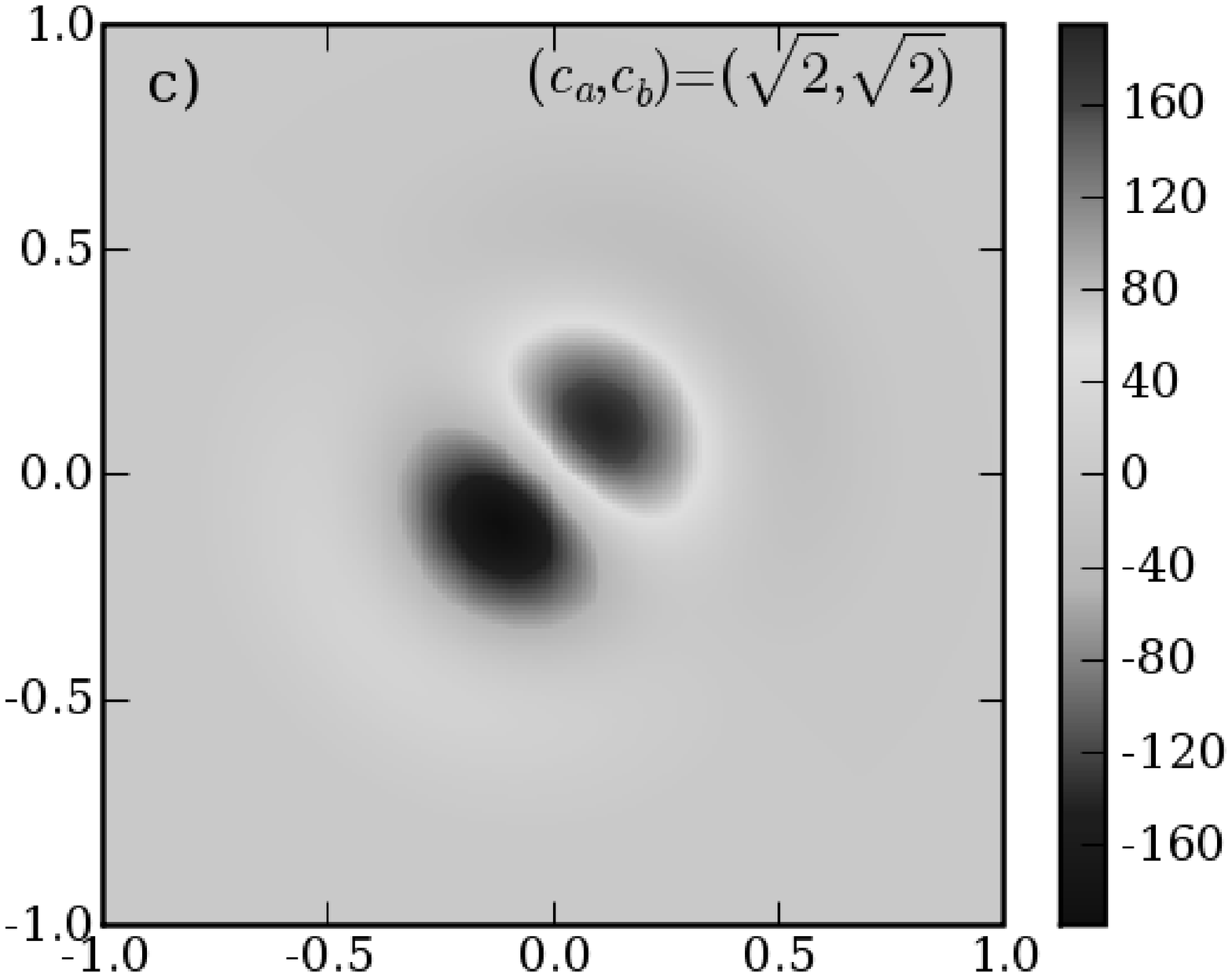}{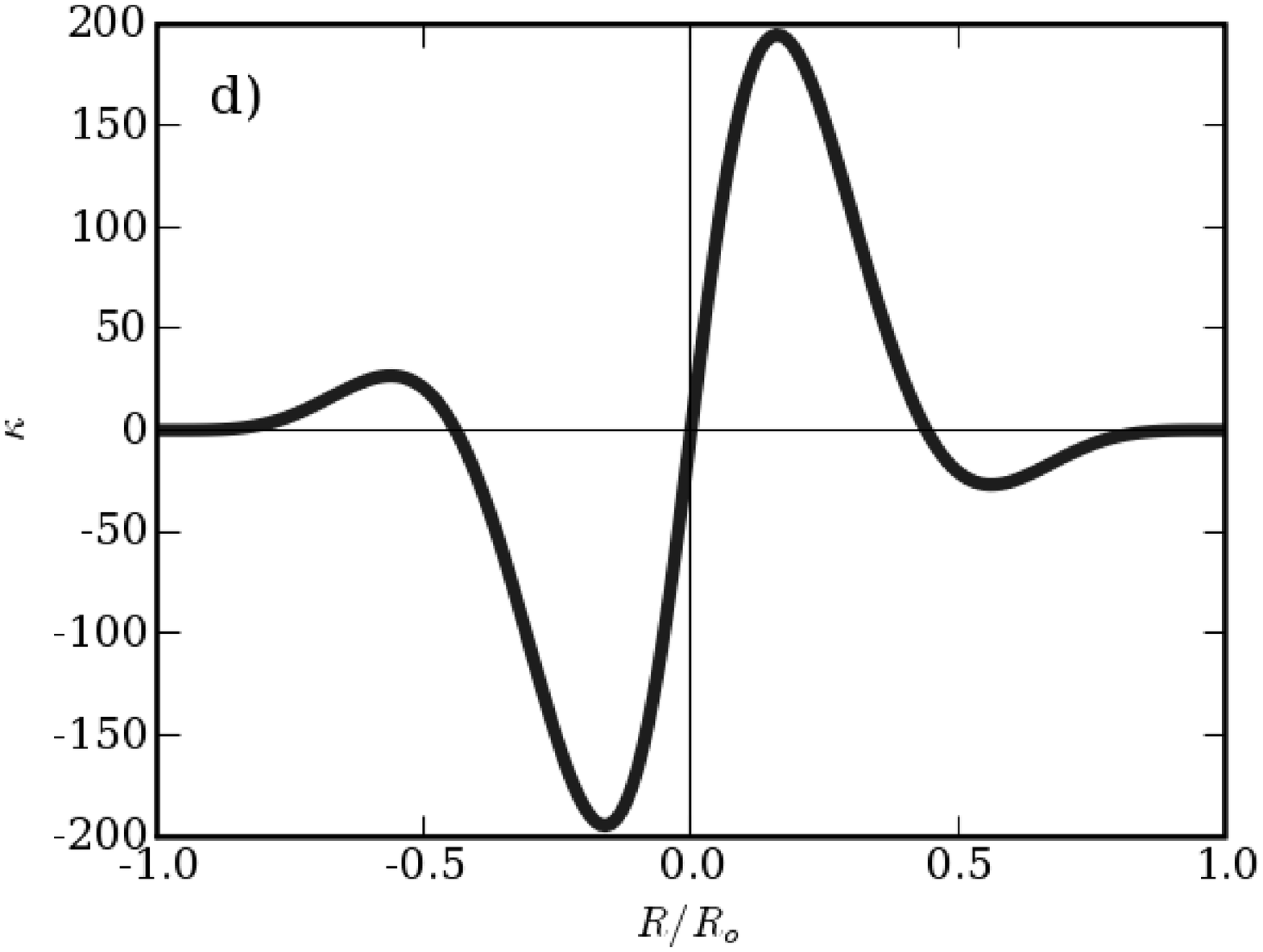}
\\
\caption[Massmap basis functions]{
\label{KBF}
Massmap basis functions constructed as derivatives of the Wendland function
$W_{7,3}$.
a) $\kappa_a$, b) $\kappa_b$,
c) $(\kappa_a + \kappa_b) / \sqrt{2}$.
d) Profile cut through the ``major axis''.
Note that outside the domain [-1, 1], the function equals zero.
}
\end{figure}

Of relevance in this work are derivatives of this function
which serve as basis functions for our massmaps.
Having solved for the deflection field
$
\left[ \alpha \right] =
\left[ \phi_{,(xy)} \right]
\left[ c \right],
$
the massmap may then be calculated analytically as:
\\
\begin{eqnarray}
\label{kappa}
\kappa & = & \onehalf \nabla \cdot \vec \alpha\\
 & = & \onehalf \sum_i^N \left[
c_{a_i} ( \phi_{,xxx_i} + \phi_{,xyy_i} ) +
c_{b_i} ( \phi_{,yxx_i} + \phi_{,yyy_i} )
\right].
\end{eqnarray}
The basis functions:
\\
\begin{eqnarray} 
\label{KBFab}
\kappa_a & = & \onehalf (\phi_{,xxx} +\phi_{,xyy})\\
\kappa_b & = & \onehalf (\phi_{,yxx} +\phi_{,yyy})
\end{eqnarray}
\\
are shown in Figure \ref{KBF}.
Different combinations of these two basis functions simply serve 
to rotate it about its axis and change its amplitude.

The basis functions, with amplitudes and orientations solved for,
are placed at the positions of the multiple images.
The resulting coefficients are generally many orders of magnitude 
greater than the amplitude of the massmap.
These large mass components all cancel out nearly perfectly with the
small ``residuals'' being the massmap solution.
Together, these mass components form a very flexible basis set.
Although the basis functions are derived from radial basis functions,
no symmetry conditions (radial, or otherwise) 
are imposed on the total massmap solution.

We note that other RBFs may be used in place of $W_{7,3}$,
including other Wendland functions, thin plate splines, multiquadrics,
power laws, and even Gaussians \citep[e.g.,][]{Fuselier06}.
Wendland functions are especially well-suited 
for use in the interpolation matrix,
but other functions may be explored in future work.

\subsection{Other Observables}

For reference, we provide here expressions for other observables
as functions of our basis function.
The lensing potential (Eq.~\ref{potentialkappa}) 
is defined as a sum of derivatives of our basis function:
\\
\begin{equation}
  \psi = c_a \phi_{,x} + c_b \phi_{,y}.
\end{equation}
Other observables may then be derived as derivatives of this potential.
The deflection field is given as the gradient of the potential:
\begin{equation}
  \vec \alpha = \psi_{,x} \hat x + \psi_{,y} \hat y.
\end{equation}
We also obtain surface mass density
\\
\begin{equation}
  \kappa = \onehalf \left ( \psi_{,xx} + \psi_{,yy} \right ),
\end{equation}
\\
shear
\\
\begin{eqnarray}
  \label{shear}
  \gamma_+     & = & \onehalf \left ( \psi_{,xx} - \psi_{,yy} \right ),\\
  \gamma_\times & = & \psi_{,xy},\\
  \gamma^2     & = & \gamma_+^2 + \gamma_\times^2,
\end{eqnarray}
\\
the inverse of the magnification
\\
\begin{eqnarray}
  \label{magnif}
  1 / \mu & = & (1 - \kappa)^2 - \gamma^2\\
  & = & 1 - \psi_{,xx} - \psi_{,yy} + \psi_{,xx}\psi_{,yy} - \psi_{,xy}^2,
\end{eqnarray}
\\
and time delays
\\
\begin{equation}  %
  \label{timedelay}
  \Delta \tau = \frac{(1 + z_L) D_S}{c D_L D_{LS}} 
  \left [
  \onehalf \left \vert \vec \theta - \vec \beta \right \vert ^2 - \psi
  \right ].
\end{equation}

\subsection{Source Positions and Massmap Physicality}
\label{sourcepos}

Image deflections are defined by the combination of 
image positions, source positions, and redshifts.
But source positions are not known in practice
and redshift measurements may contain large uncertainties.
In our method, these become our free parameters.
Each set of source positions and redshifts
will redefine the deflection field,
yielding a different LensPerfect massmap solution.
We iterate to find the most plausible solutions
including a single ``best'' solution, as we describe below.
An ideal optimization procedure would lead us directly
to the true source positions and redshifts.
The optimization procedure we have developed may not be ideal,
but we believe it produces 
an accurate ``best'' solution (see \S\ \ref{A1689model}),
similar to that obtained with the true source positions.
We describe our optimization procedure in the next subsection 
(\S\ \ref{optimization})
after first describing our method for rating solutions.

LensPerfect will return a perfect massmap solution 
for any set of input source positions and redshifts.
But only certain sets will yield physical solutions:
those with positive mass everywhere within the convex hull,
where our solutions are constrained.
And some sets will yield ``more physical'' solutions than others.
An extreme example of a ``less physical'' solution
would be a ``donut'' solution
comprised of a high density ring surrounding a lower density center.
By developing a method to rate different solutions as more or less physical,
we can iterate over possible source positions to find 
those which yield plausible solutions,
including the most plausible or ``most physical'' massmap solution.
Our goal is to discard unreasonable solutions without biasing our result 
toward our concept of what a massmap should look like
(that it should have an NFW profile, for example).

In the PixeLens method, a series of rigid criteria are used to distinguish
``physical'' from ``unphysical'' massmaps \citep{SahaWilliams04}.
They define a physical massmap as one that is positive everywhere
and satisfies restrictions on 
maximum pixel-to-pixel variation and direction of the massmap gradient.
Additional constraints are optionally imposed:
inversion symmetry about the axis and a minimum radial slope.
And early incarnations of PixeLens found those massmap solutions
which most closely followed the light distribution
\citep[as advocated in][]{SahaWilliams97, AbdelSalam98a}.

After some experimentation, we have developed a new measure of physicality.
Rather than imposing rigid arbitrary constraints, 
we assign a figure of merit to each massmap solution
based on the following physicality traits:

\begin{enumerate}
\item Positive mass everywhere within the convex hull
\item Low mass scatter in each radial bin,
thus preferring azimuthal symmetry
\item No ``tunnels'': penalty for individual mass pixels 
within the convex hull
being ``too low'' relative to others at similar radius
\item Overall smoothness: minimal pixel-to-pixel variation 
within the convex hull
\item Average mass in radial bins decreases outwards
(penalty for increasing outward)
\end{enumerate}

Below, we motivate this physicality list
which has been chosen for the purposes of this paper.  
Note the user has the ability to modify
this list rather straightforwardly 
given the freely-available LensPerfect software.

The first trait is our only rigid constraint,
obviously required of any physical massmap.
Note that our massmaps are only constrained within the convex hull.
Our basis functions do necessarily yield negative mass outside this region,
but this is not necessarily a concern.
We note that we are able to ``get the right answer'' 
inside the convex hull (Figs.~\ref{A1689kLP}, \ref{A1689kLPopt}).
And if negative mass lies in
a symmetric ring outside the convex hull,
then it will have no effect on the image positions inside.

What we would like to avoid is one or more large isolated pockets 
of negative mass off to one side outside the convex hull.
These may arise as ``corrections'' to a solution which is not quite accurate
within the convex hull.
Large positive mass clumps outside the convex hull are similarly undesirable.
Our second physicality trait serves to beat down these clumps,
along with other positive mass clumps and underdense pockets
within the convex hull.
We may worry that this biases against the presence of subhalos.
But turning this around, if we are biasing against subhalos,
then we can be more confident of any subhalos
that do arise in our solutions.
And Occam's razor would dictate that
the massmap solution with the fewest subhalos is most likely.
Also note we are talking about subhalos in large sub-clumps.
Smaller satellites generally cannot be resolved.
(Our resolution is limited by the density of multiple images 
observed in the lens plane.)
In the future, we may wish to perform tests
to determine exactly how well LensPerfect
recovers different amounts of substructure in large subhalos,
given different numbers of multiple images.

Underdense pockets incur extra penalty via our third physicality trait.
While a small overdense area may be due to substructure,
a small underdense area is much less physical.
As our massmap is integrated a long the line of sight,
an underdense pocket suggests that 
a ``tunnel'' has been carved through the entire mass distribution.
This is not very plausible, and thus
we penalize such tunnels more severely than overdensities.

We further beat lumps and pockets out of our solutions
by minimizing pixel-to-pixel variation of mass
(our fourth physicality trait).

Finally, we penalize (without forbidding) mass profiles 
which increase with radius.
Of course this might bias against spectacular findings
like the ring-like structure in CL0024 reported by \cite{Jee07}.
But we believe that it is perfectly healthy 
to bias against such spectacular solutions.
If a more pedestrian solution may be found which exhibits no ring structure,
then it is likely that no ring structure exists.
And if such a ring-like structure were to persist in our solution 
despite this bias, we could be that much more confident of its existence.
Additional tests could then be performed to show how well LensPerfect
recovers ring structures in known massmaps
and whether it might ``recover'' rings when they are not present.
Finally, we note that in CL0024, the ring-like structure is detected 
(and predicted in simulations of a cluster collision)
to lie well outside the strong lensing region.

Note that our physicality traits 2, 3, and 5
assume at least a rough azimuthal symmetry.
We have yet to experiment with mass distributions which are strongly 
multi-modal.
While our current method will suffice for many clusters, including Abell 1689,
we will probably need to revise our penalty functions
to analyze a highly asymmetric cluster such as Abell 2218.

Even having settled on a set of physicality traits,
it is unclear how exactly to calculate penalty functions based on them.
And once these penalty functions are established,
how should they be weighted relative to each other?
The goal is for equally ``offensive'' massmaps to be penalized equally.
Again this is highly subjective, but after much experimentation,
we have devised a total penalty function which works well.
The details are given in the Appendix \ref{penalty}.

\subsection{Massmap Optimization}
\label{optimization}

With a good penalty function in hand, we may then proceed
to find those source positions and redshifts
which produce the most physical massmap. 
Given many lensed galaxies, this is far from trivial. 
For example, 19 source positions ($x$, $y$) yield 38 parameters 
which must be optimized over,
plus up to 19 redshifts with their associated uncertainties.
And for each iteration, 
we must obtain a low resolution massmap to calculate our penalty function.
This takes from less than a second to a few seconds as more images are added.
Thus we must choose a scheme which efficiently navigates
our 38(+19)-dimensional parameter space, 
minimizing the number of iterations necessary.
Fortunately, we may use a few tricks to converge to good solutions 
fairly quickly.

The first trick is commonplace 
in strong lensing massmap reconstruction methods:
we build the massmap one galaxy at a time.
(That is, we add the multiple images of each galaxy in turn
as constraints to our model.)
But we take this a step further,
tearing our solution down and rebuilding it for every iteration.
These rebuilds are extremely quick, 
as at each step, the deflection field solution 
need only be calculated at the new image positions.

\begin{figure}
\plottwo{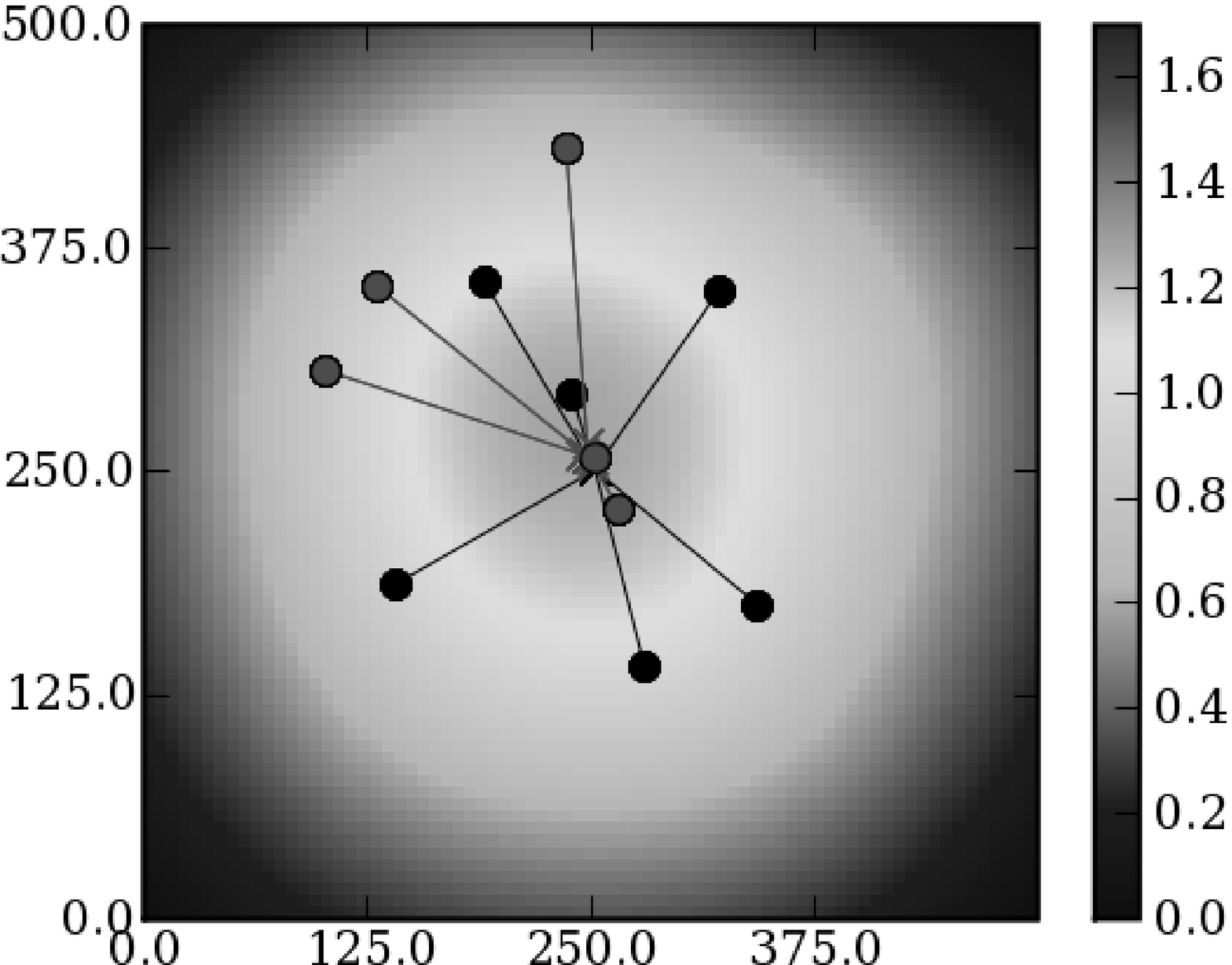}{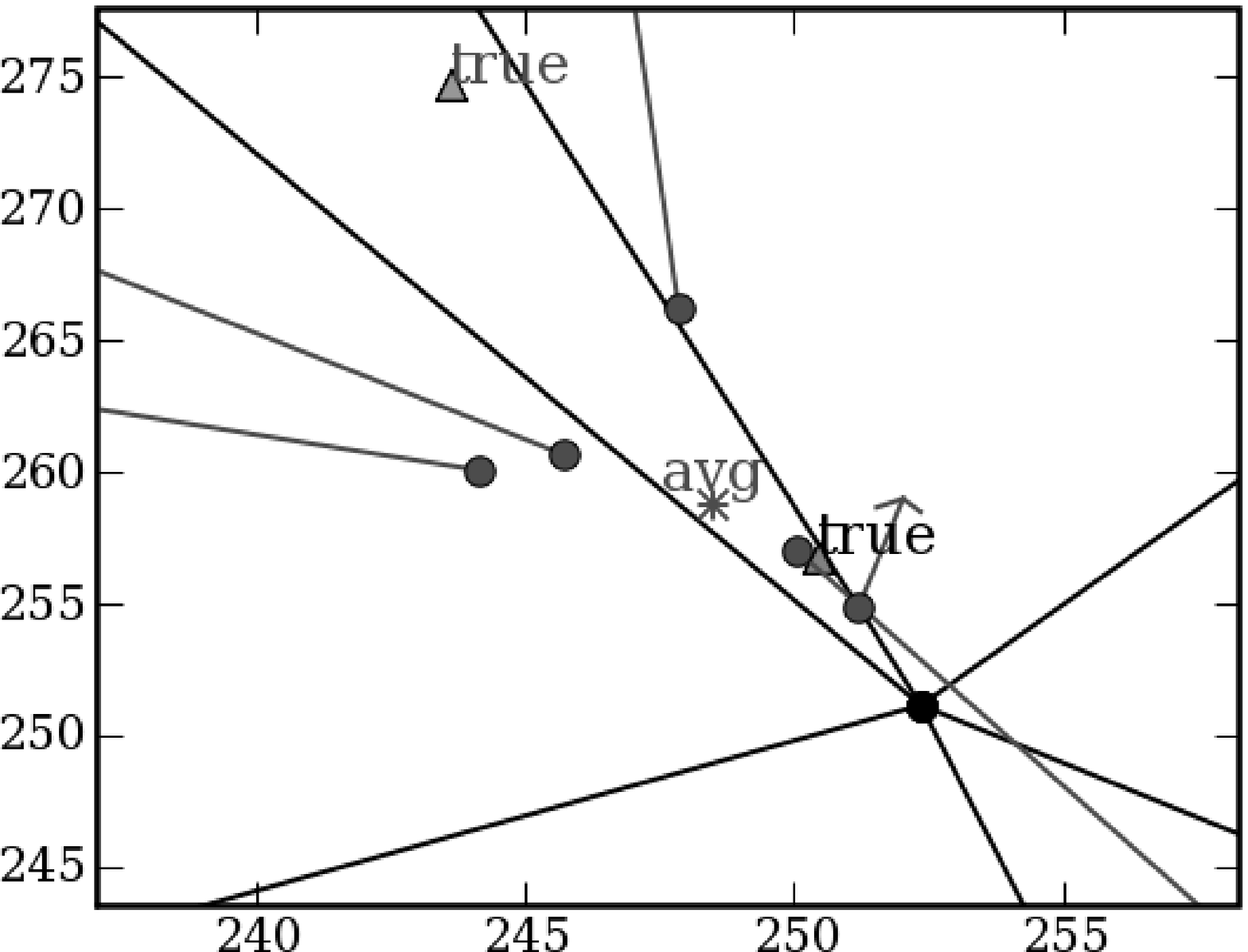}
\plottwo{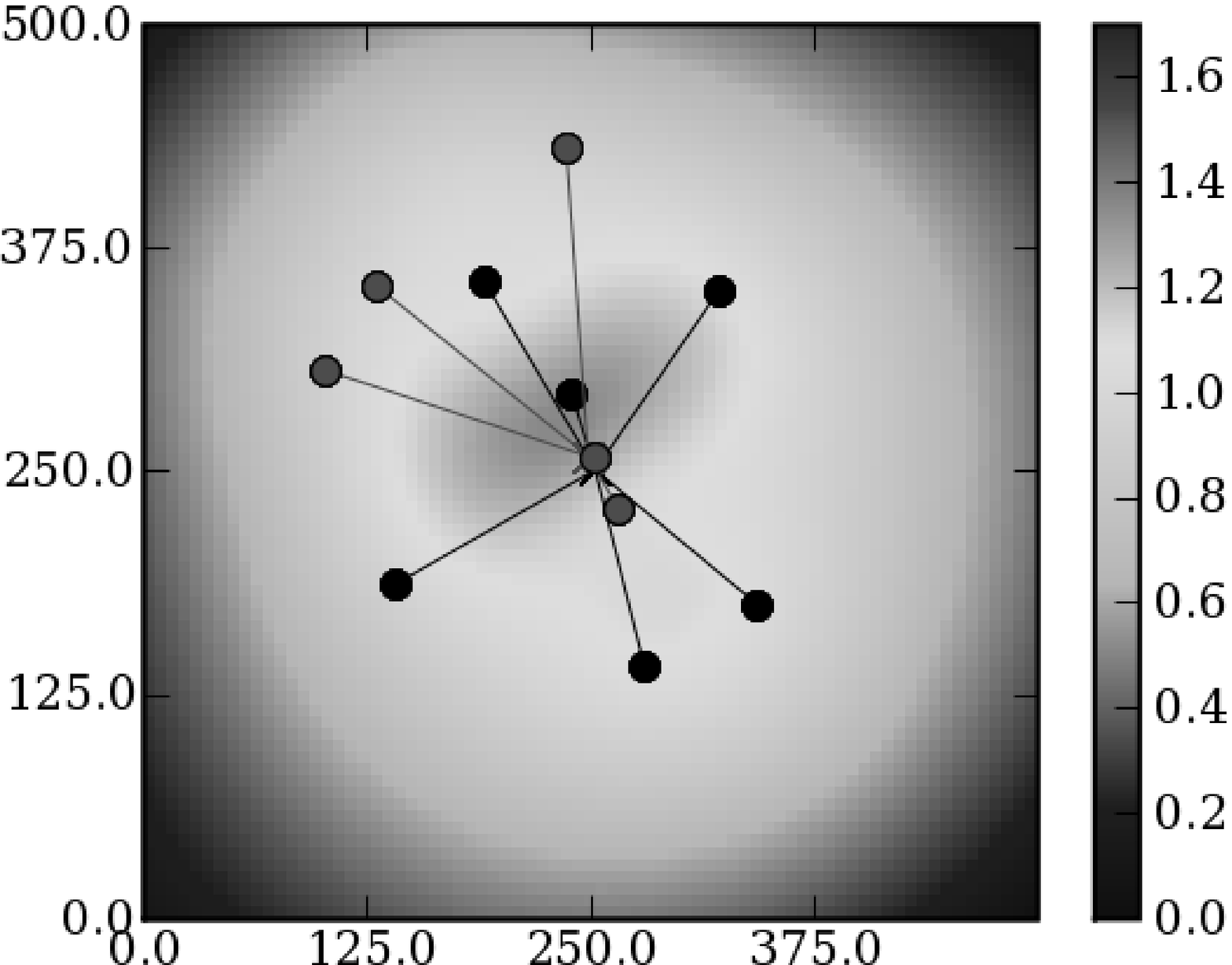}{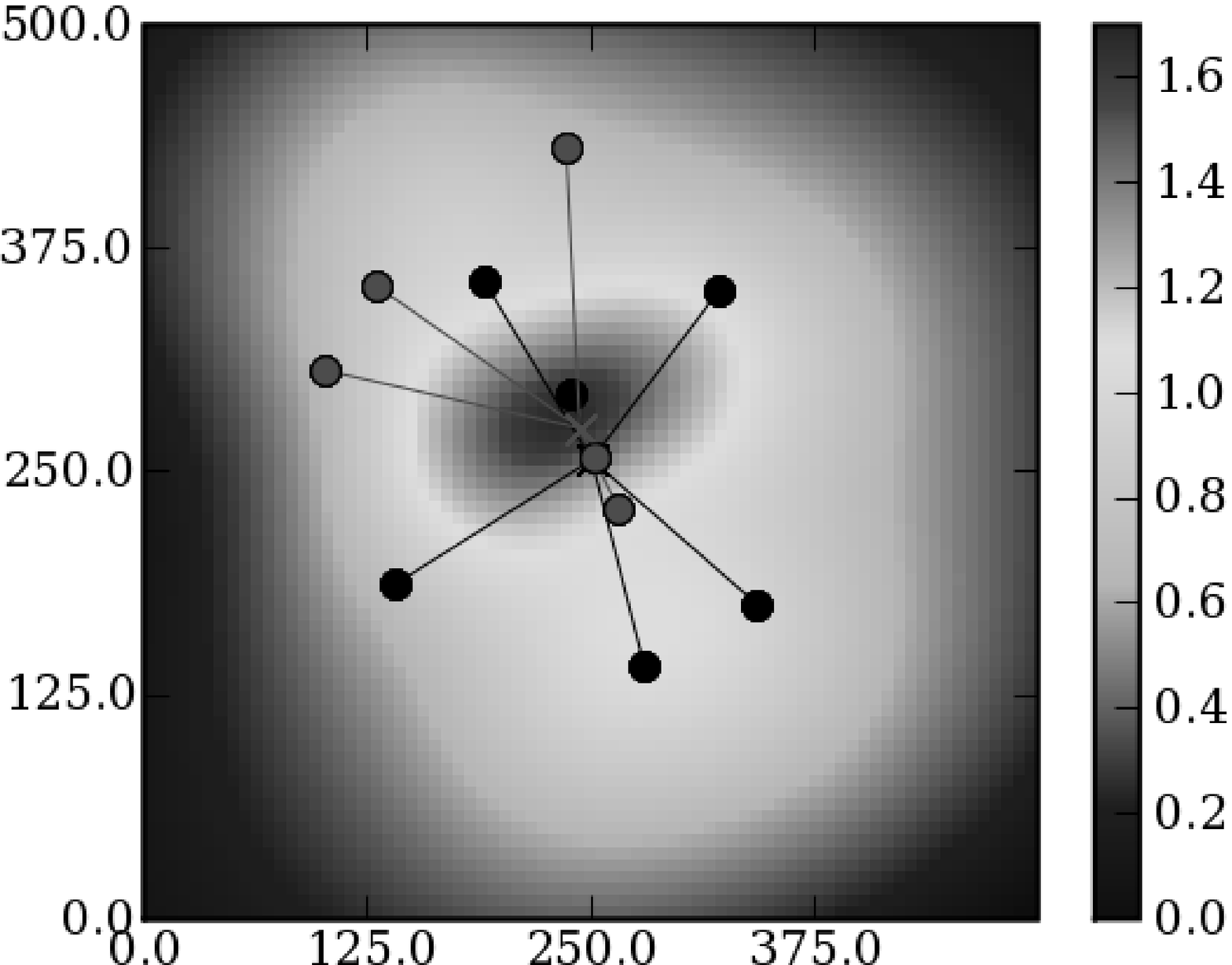}
\caption[Source position derivation]{
\label{source2}
Upper left: Massmap obtained using a single lensed galaxy (black):
object \#1 in our mock A1689 model.
Based on this solution, a galaxy \#2 (red)
is delensed back to the source plane.
As we see in the enlarged plot (top right)
the de-lensed positions (now plotted as circles) do not exactly align.
We take the (weighted) average of these source positions,
labeled with the (*) and ``avg'',
and assign this to the source position for this galaxy.
Also plotted (triangles) are the true source positions
for objects \#1 and \#2.
Bottom left:
Massmap solution obtained once both source positions have been assigned
as above.
By perturbing these source positions, we obtain a range of solutions
all of which perfectly reproduce the observed image positions.
Bottom right:
The solution obtained
when both sources are assigned to their true model positions.
Strictly speaking, this is a mock model so the scale is arbitrary.  
But if this were actually A1689, the scale would be $0\farcs 40$ / pixel.
}
\end{figure}

As we add each galaxy to our model, we place its source position
using the average of those predicted from the current solution
(Fig.~\ref{source2}).
Each image will delens back to a slightly different position
(as these images have not yet been constrained).
By taking the average of these (weighted, as discussed below),
we can obtain a good initial guess as to its source position.

Given this source position we will obtain a new massmap solution.
But we can also perturb this source position
and obtain a different massmap solution instead.
Thus we iterate 
and search for that source position which yields the most physical solution
according to our penalty function.
We perform this 2-dimensional optimization
using the \cite{Powell64} routine\footnote{More elaborate
routines are available, but these may actually be less efficient.
Specifically, calculating gradients of the penalty in source position space
would require two extra and time-consuming function evaluations 
for every iteration.}
included in the SciPy Python package.\footnote{http://www.scipy.org}
If the redshift of this galaxy is uncertain, 
we may also optimize the redshift at this time
(including a penalty if it drifts too far from its expected value).
(For this one-dimensional optimization, 
we use the simple golden section search method,
also included in SciPy.)
Once the source position and redshift have been optimized, 
we may proceed to adding the next galaxy.

Before proceeding to the next galaxy, 
we may choose to re-optimize all previous source positions and redshifts.
This ``reshuffling'' can improve the overall solution.
But once, say, 10 galaxies have been placed,
we may worry that our solution has been ``locked in''.
Attempting to re-optimize the third galaxy position
will not be very effective,
as it is now tightly constrained by the positions of the other nine galaxies.

Thus we use a flexible parameterization
in which the perturbations to the predicted source positions
(as described above) constitute our free parameters to be optimized.
That is, we maintain a list of source position offsets 
(one for each galaxy, initially set to $(0,0)$),
and it is this list which we optimize,
rather than the source positions themselves.
Every time we rebuild our massmap solution, 
we obtain each source position as shown in Fig.~\ref{source2},
but then we offset it according to our list of offsets
before placing it and proceeding to add the next galaxy.
Note that each offset affects not only the source position of that galaxy,
but also (as this modifies the solution) those of all galaxies that follow.
In essence, source positions move and adjust with each other.

As mentioned above, we determine each new source position
by taking {\it weighted} averages of those predicted.
We assign more weight to a delensed position
for which the image is close to an already established image.
For example, in Fig.~\ref{source2}, a red image near any black image position
will have its source position given more weight in the average.
The idea is that if the deflection field has already been established 
at a (black) image position,
then the deflection field at the nearby (red) point should be similar.
We have found this weighting scheme yields better source positions
(closer to the true model positions)
than straight averaging.
We also had some success giving more weight
to sources with images at large radii
(where there is less mass available to support rapid changes in the 
deflection field).
In the end, the exact scheme is somewhat irrelevant
as offsets from these average positions will be perturbed and optimized.
But it is possible that a better scheme would yield quicker convergence.

Above we have described all the details of our optimization scheme,
except for how we begin!
That is, where do we put our first source position?
Before any galaxies are placed, we have no massmap solution,
and thus no predictions for the first source position.
Fortunately just about any choice will do for the first source position, 
as all yield identical (or nearly so) solutions within the convex hull. 
Even after additional sources are added, shifting the first source position 
has little effect on the overall solution.
The entire source plane basically shifts along with it,
yielding a nearly identical solution within the convex hull.
This is a well-known lensing degeneracy,
but certain shifts in the source plane
do yield more physical solutions than others,
as we demonstrate in \S\ \ref{massstick}.
Correct source positions yield a solution
which is more symmetric outside the convex hull.

\subsection{Constraining Image Fluxes and/or Shears}
\label{fluxes}

Image fluxes may provide additional constraints to the mass model.
To constrain the fluxes (and shears) of lensed images,
previous authors have added terms to their matrix equation
relating source positions, lens parameters, and observables 
\citep{EvansWitt03,CongdonKeeton05,Diego07}.
However we prefer not to interfere with our Eq.~\ref{deflmatrix}
which, given proper basis functions, is guaranteed to obtain a perfect
interpolated solution of any input deflection vectors.
Fortunately, relative image fluxes (magnifications) and shears
are determined by local rates of change in the deflection field and thus
may be easily and precisely constrained by adding deflection field constraints.

\begin{figure}  
\plotone{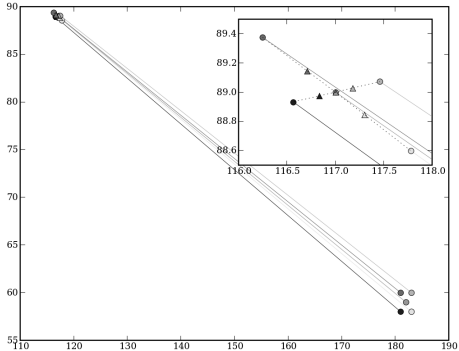}
\caption[Flux constraints]{
\label{fluxcontrol}
Method used to constrain image fluxes.
A small box is constructed around each image (bottom-right),
and it is delensed back to the source plane (top-left)
using the initial massmap solution
(that derived using the image positions only).
The inset zooms in on the delensed positions (circles).
The ratio of the area of this parallelogram 
to the ratio of the ``box'' in the source plane
yields the magnification of this image.
We then adjust this ratio by adjusting the size of the delensed image
(modifying our deflection field at the corners of the box).
In this case we require a larger magnification,
so we decrease the size of the source image
to that defined by the triangles.
Our image box is constrained to delens to this smaller source,
thus adjusting its magnification.
In practice we find that rather than 
adding four constraints to the deflection field,
we need only constrain two points, 
one above or below and the other to the left or right of the image.
}
\end{figure}

In this subsection we describe how to add flux and/or shear constraints
for {\em multiple images}.
(Also see \S\ \ref{knots}.)
We do not yet have a method for incorporating weak lensing shear
or any other constraints from singly-imaged galaxies
(though see \S\ \ref{weaklensing}).

To constrain the fluxes of multiply-imaged galaxies,
we begin by obtaining an initial massmap 
that reproduces the image positions.
Next, we construct a small box around each multiple image
and delens each back to the source plane.
We measure the area of each delensed box
(now a parallelogram in the source plane)
and compare to its original area in the image plane.
The ratio of these areas yields the magnification for that image,
at least as predicted for our initial model.\footnote{Of course
we can also calculate the magnification using Eq.~\ref{magnif}.}
Now, by modifying the deflection field at the corners of the box,
we can adjust the relative sizes of the lensed and unlensed boxes 
so that they produce the proper
(observed) magnifications (Fig.~\ref{fluxcontrol}).
Given these new deflection field inputs,
which encode both position and flux constraints for the multiple images,
we obtain a new massmap solution 
that perfectly reproduces the observed positions and fluxes.

We note that image shears could be constrained in a similar manner, if desired,
by adjusting the relative shapes of the lensed and delensed boxes.
And observational uncertainties may be incorporated
by adding multiple realizations of noise to the measurements
and finding a perfect solution to each,
as done by \cite{EvansWitt03} and \cite{CongdonKeeton05}.

In practice we find that rather than 
adding four constraints to the deflection field for each flux measurement,
we need only constrain two extra points, 
one above or below and the other to the left or right of the image.
This not only helps reduce computing time,
but it also keeps our number of free parameters 
more in line with the number of observable constraints.
We could keep the two numbers exactly equal by adding a single constraint.
But this fails to properly constrain the flux,
instead squeezing mass out to the sides as when one sits on a balloon.

\subsection{Computational Efficiency}
\label{compu}

Other strong lensing analysis methods (with the exception of PixeLens)
face a dilemma over whether to minimize scatter
in the source or image plane.
The former choice, while less robust and subject to possible biases,
is often chosen for computational efficiency.
LensPerfect does not have to make this choice
as both source and image positions are always perfectly constrained.

And LensPerfect is computationally efficient.
Once all source positions and redshifts have been established
(along with image positions),
our massmap solution coefficients are obtained
``instantly'' (in a fraction of a second)
via direct matrix inversion without need for iterations.
Evaluating this massmap solution on a grid,
while not quite ``instant'', is still very fast.
On a Mac Powerbook G4 laptop, given $N_i = 93$ multiple images,
the massmap can be evaluated on a $N_p = 2500 = 50 \times 50$ grid
in 3 seconds, scaling with $N_i N_p$.

\section{Applications}
\label{sec:applications}

\subsection{Massmap Recovery Test}
\label{A1689model}
\begin{figure}
\plotone{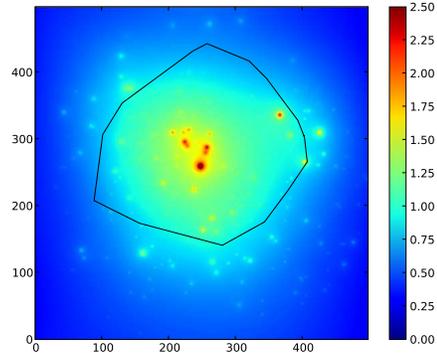}
\caption[Input model massmap]{
\label{A1689kB05}
Input massmap to be recovered with LensPerfect given mock lensing data.
This is one of the early massmap solutions of Abell 1689 
obtained by \citet{Broadhurst05}.
Mass is plotted in units of critical density for a source at redshift infinity.
The black line reappears in Fig.~\ref{multimages}
where its meaning is readily apparent.
}
\end{figure}

\begin{figure}
\plotone{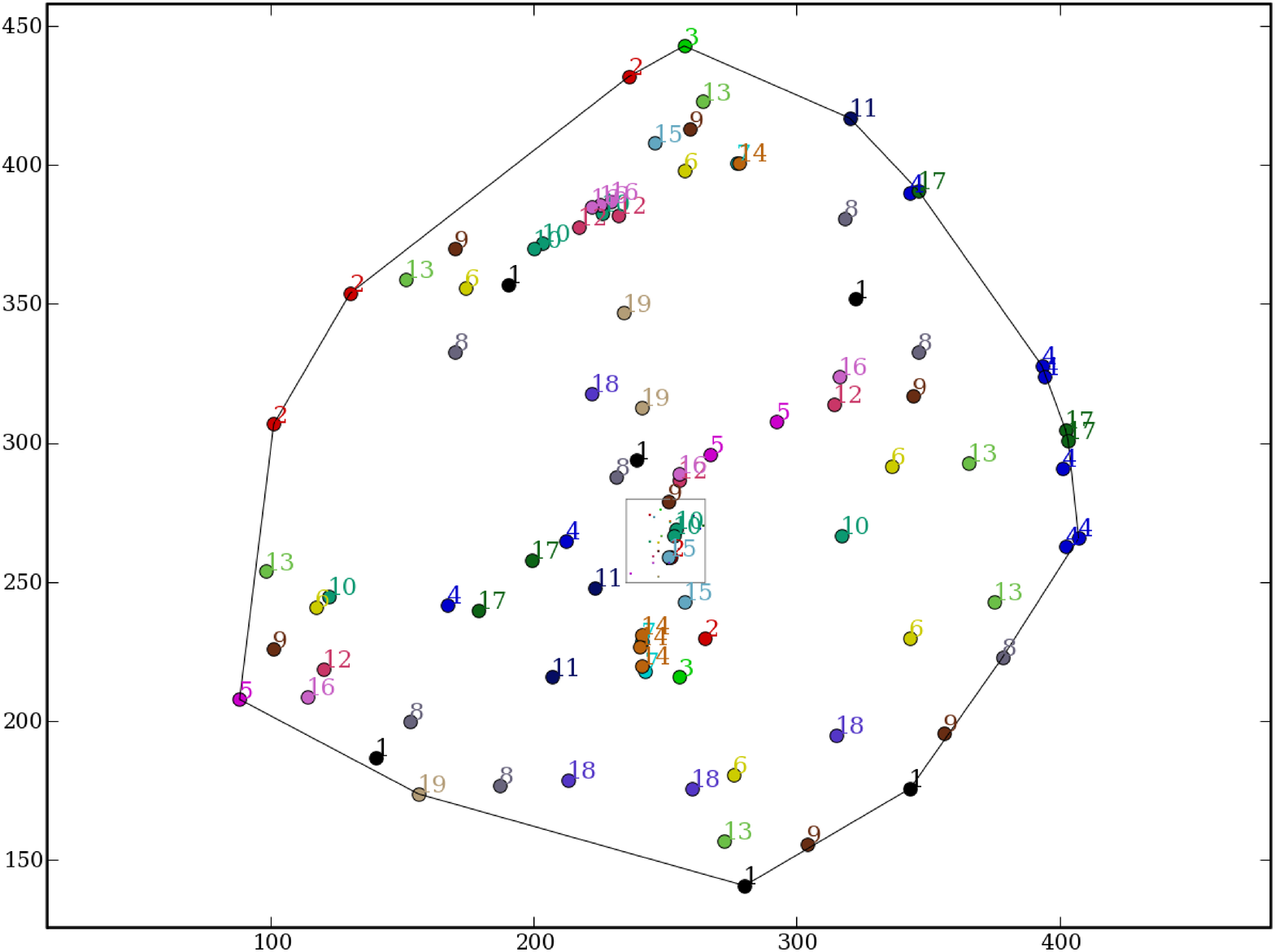}
\caption[93 multiple images of 19 model galaxies]{
\label{multimages}
The input massmap (Fig.~\ref{A1689kB05}) 
is used to lens 19 model galaxies
to produce 93 multiple images at the positions shown here.
Each image is colored and labeled according to its source galaxy.
The source galaxies are plotted as small dots framed by a gray square.
The black line is the ``convex hull'' which traces the outermost images.
Inside this region, the deflection field is interpolated
and thus our solution is well constrained.
Outside, the extrapolated deflection field 
allows a much wider range of solutions.
}
\end{figure}

\begin{figure}
\plotone{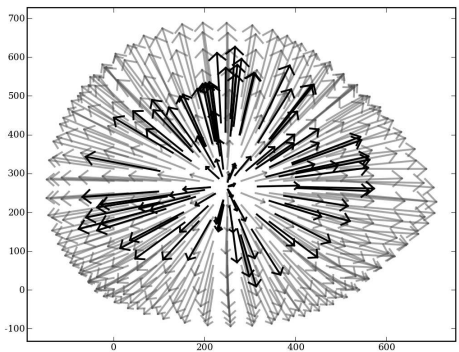}
\caption[Interpolated deflection field]{
\label{A1689defl}
Black: Deflections of the 93 images in Fig.~\ref{multimages}.
All deflections are normalized to a source redshift of infinity,
and the tail of each vector has been moved outward
to the corresponding image position.
Red: LensPerfect interpolation of the deflection field
with $R_o = 700$ in the units plotted.
The interpolation is curl-free 
and exactly matches the vectors at the given data points.
}
\end{figure}

\begin{figure}
\plotone{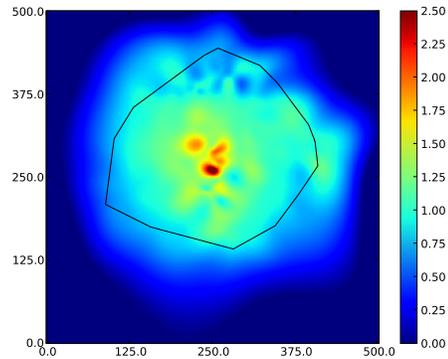}
\caption[Recovered massmap solution, given source positions]{
\label{A1689kLP}
LensPerfect massmap solution given the 93 images in 
Fig.~\ref{multimages} with known source positions and redshifts
and setting $R_o = 700$ in the units plotted.
The massmap is simply half the divergence of the deflection field
plotted in Fig.~\ref{A1689defl}.
The scale is identical to Fig.~\ref{A1689kB05}, the input massmap.
What we obtain with LensPerfect given 93 images
is basically a low-resolution interpolation of the input massmap.
Note that the significant features 
are reproduced inside the convex hull (the black line).
Outside this line, the solution is very poorly constrained,
and can be varied simply by changing $R_o$.
Here, the mass outside goes a bit negative 
(we have neatly clipped this from our colormap).
But when we increase $R_o$ to 1500,
we find the mass is positive within the entire region plotted,
while the massmap inside is barely affected.
}
\end{figure}

\begin{figure}
\plotone{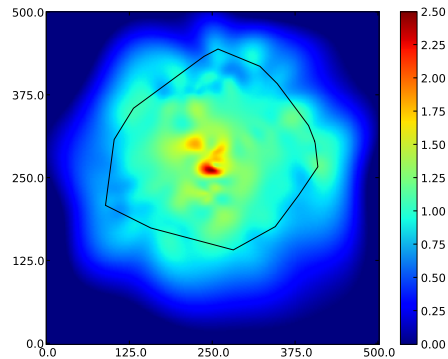}
\caption[Recovered massmap solution with optimized source positions]{
\label{A1689kLPopt}
LensPerfect massmap solution given the 93 images in 
Fig.~\ref{multimages} with known redshifts
but without knowledge of the input source positions.
We found optimal source positions,
or those which produced the most physical massmap,
as judged by the criteria we describe in the text.
}
\end{figure}

Here we demonstrate our technique given a mock galaxy cluster 
with simulated gravitational lensing.
More important than the ability to perfectly reproduce 
all multiple image positions
is the accuracy to which we can recover the true massmap distribution.

We would like to test LensPerfect for a case similar to that observed 
in Abell 1689.
Thus we create a mock galaxy cluster ``Babell 1689'',
which is very similar to Abell 1689.
In fact, the massmap of Babell 1689 (Fig.~\ref{A1689kB05})
is actually a solution obtained by
\cite{Broadhurst05} from their analysis of Abell 1689.
But for our purposes, Babell 1689 is just a mock galaxy cluster.
We use it to gravitationally lens 
19 mock galaxies at redshifts between 1 and 5.5, 
producing 93 multiple images (Fig.~\ref{multimages}).
This is similar to the number of multiple images (106) identified by
\cite{Broadhurst05}.

We stress that we are not analyzing Abell 1689 here.
Our analysis of Abell 1689 and its multiple images
will be published in an upcoming paper.
Here we are analyzing the mock cluster Babell 1689 
given its mock multiple images.

The 93 mock multiple image positions 
along with 19 source positions and redshifts
are fed into LensPerfect as input.
Fig.~\ref{A1689defl} shows
the input deflection field scaled to a source redshift of infinity
along with a LensPerfect curl-free interpolation.
The solution was obtained using the
Wendland function given in Equation \ref{WendlandC6}
with a scale factor of $R_o = 700$ in the units plotted.
One half the divergence of this deflection field gives 
the LensPerfect massmap solution (Fig.~\ref{A1689kLP}).
Note that it is basically a low-resolution interpolation of the input massmap
(Fig.~\ref{A1689kB05}).
No assumptions were required about the form of the massmap,
and yet the general form and significant features 
of the input massmap are recovered.
Were more than 93 input images given, 
the resolution would increase and finer details would be resolved 
(\S\ \ref{1000}).

Note that the \citet{Broadhurst05} solution 
may appear to resolve very fine detail in the Abell 1689 massmap
absent from our reconstruction of Babell 1689 presented here.
But this is just a result of their assumption 
that some component of the Dark Matter traces light.
LensPerfect makes no such assumption.

In practice, we will not have knowledge of the source positions.
So we repeat our analysis, but without providing the source positions as input.
Instead we use the source position optimization method
outlined in \S\S\ \ref{sourcepos} and \ref{optimization}
and detailed in the Appendix \ref{penalty}.
The ``most physical'' massmap solution we find
is shown in Fig.~\ref{A1689kLPopt}.
It is very similar to that obtained with the true source positions
(Fig.~\ref{A1689kLP}).
In this case the only information we input to LensPerfect
were the image positions and redshifts, 
which we assumed had been measured with perfect precision.\footnote{Redshift 
uncertainties vary greatly from one study to the next.
Thus rather than attempting to present a test which includes
``typical'' redshift uncertainties,
we propose that such tests be performed on a case-by-case basis.
We note that we have had success in modeling real-life data
such as the actual Abell 1689 multiple images
complete with their redshift uncertainties
(Coe et al.,~in prep.).}

\begin{figure}
\plotone{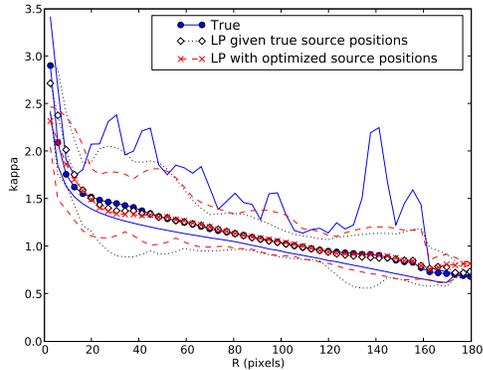}
\caption[Radial profile comparison]{
\label{profile}
Radial profiles of model (Fig.~\ref{A1689kB05}) 
and recovered massmaps (Figs.~\ref{A1689kLP} and \ref{A1689kLPopt}).
Points within the convex hull are binned in radius,
and the mean $\kappa$ are plotted as symbols 
with lines demarking the minima and maxima.
}
\end{figure}

Radial profiles of the three massmaps 
(model, recovered with true source positions, 
and recovered with optimized source positions)
are compared in Fig.~\ref{profile}.
Only points within the convex hull are considered.
Agreement of the mean mass is excellent 
except for some deviation in the center
and slight deviations toward the outside of the convex hull.
The fine structure of the mass peaks is not recovered
in these modest-resolution (93 multiple image) massmaps.

\subsection{Source Position Recovery}
\label{sourcepossec}

How well do our optimized source positions match the true source positions?
Inaccuracies in the source positions propagate to the deflection field 
and thus to our massmap.
Modest shifts of the entire source plane are acceptable, however.
This well-known degeneracy does not affect the solution inside the convex hull
(see \S\ \ref{massstick}).

\begin{figure}
\plotone{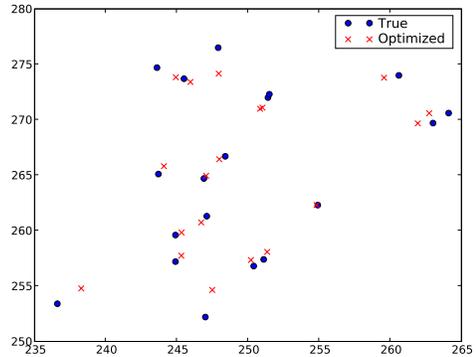}
\caption[Optimized vs.~true source positions]{
\label{sourceposplot}
Optimized source positions compared to the true input source positions.
All of the optimized positions have been offset by (+2.1, -6.2)
to bring them in line (on average) with the true positions.
(The solution is very insensitive to such shifts, 
especially within the convex hull.)
The optimized source positions are slightly biased 
toward a solution with higher magnification.
The solution they yield (Fig.~\ref{A1689kLPopt}) is perfectly valid.
And it is a simple exercise 
for the user to spread the source positions further apart
and explore solutions with lower magnifications.
}
\end{figure}

With this in mind, we plot (Fig.~\ref{sourceposplot}) 
our optimized and true source positions from the previous subsection.
A constant shift of a few pixels has been added to the optimized positions
to bring them in line (on average) with the true positions.
This is justified (not just for our method, but for any method)
as shifts in the source plane are a degeneracy in the problem.
A massmap solution obtained with one shift is 
virtually identical and no less accurate inside the convex hull 
than a solution obtained with a different shift (\S\ \ref{massstick}).

Applying this shift,
we find our recovered source positions are offset from the true source positions
at the rate of 1.05 pixels on average for the 19 systems.
If this were A1689, 1.05 pixels would translate to $0\farcs42$.
But this value cannot be directly compared to any published results for A1689
(or any other cluster).
Source position offsets are likely inherent to all methods,
but they can never be measured in practice
since the true source positions are unknown.
The only measurable quantity is the scatter of 
(delensed) source positions within each multiple image system.
For LensPerfect, this scatter is zero.
In other methods, the errors due to scatter and offsets
may be cumulative.

Our optimal source positions are also 
biased toward a solution with slightly greater magnification
than the true solution.
This is most likely a consequence of our physicality measure
which rewards smoother massmaps.
A smoother massmap will have a shallower profile and thus higher magnification.
While we have made every effort not to bias profile slope,
some small bias may still remain.
We will work to reduce both the offsets and this bias in the future.
In the meantime, it is a simple exercise for the user to
spread out the optimal source positions
and explore solutions with lower magnifications.
In fact, this should be a part of any comprehensive analysis.

We must keep in mind the ultimate goal in our analysis.
The goal is not to obtain the single best massmap,
but rather to determine the range of massmaps 
which produce reasonable solutions.
Both of the massmaps presented in \S\ \ref{A1689model}
are perfectly valid solutions 
which perfectly reproduce the 93 multiple image positions.
We cannot know which is more accurate without knowledge of the source positions.
And of course this knowledge is unattainable in practice.
This is just an inherent degeneracy in the problem.
We are developing methods to thoroughly explore the solution space
and will report on them in future work.

\subsection{1,000 Multiple Images}
\label{1000}

The resolution of a LensPerfect massmap is dictated by 
the density of multiple images detected.
Each multiple image samples the deflection field at a given location.
In the gaps between these images, the deflection field must be interpolated,
and the exact form of the massmap becomes less certain.

\begin{figure}
\plotone{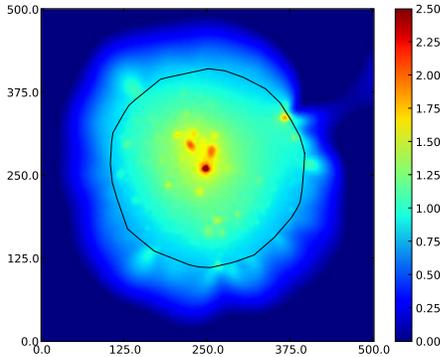}
\caption[1,000 points of light]{
\label{kappa1000}
Massmap solution to that shown in Fig.~\ref{A1689kB05}
given the deflection field sampled at 1,000 points
within the area outlined in black.
Fine detail in the input massmap is faithfully reproduced.
While this demonstration is idealized,
we may expect a massmap of similar quality 
given a lens which produces 1,000 multiple images.
}
\end{figure}

To demonstrate this, we consider the massmap solution we may obtain given
1,000 multiple images.
Rather than performing mock lensing to produce multiple images
as in the previous subsection,
here we will simply sample the deflection field at 1,000 points.
We restrict these samples to an area of similar size as before
(a circle of radius 150 pixels).
And we ensure that all samples are separated by 2 pixels or more.
Given these 1,000 samples, 
we then interpolate to find the deflection field elsewhere.
Our massmap solution is shown in Fig.~\ref{kappa1000}.
Comparing with the input massmap (Fig.~\ref{A1689kB05}),
we can see that very fine detail is faithfully resolved.

This setup is a bit idealized.
The ``multiple images'' are spread fairly uniformly (randomly)
about a circle within the field.
In practice we are more likely to find clumps and voids of multiple images
due to the magnification pattern,
obscuring foreground galaxies, 
and physically linked background galaxies.
Also, we have assumed the equivalent to
perfect knowledge of the source positions.
Nevertheless, this massmap gives us a rough idea of the resolution 
we may expect given observations deep enough to reveal 1,000 multiple images.

In practice, we would need to find and optimize 
a large number of source positions.
1,000 multiple images may be produced by about 300 background galaxies.
This saddles us with 600 free parameters.
But even in our 93-image system, we find that source positions
added toward the end are already fairly well constrained 
by those added previously.
So we can speculate that the next 200+ source positions
may similarly ``fall into place'',
making the computational challenge more manageable.

\subsection{Fewer Constraints and the Role of $R_o$}
\label{Einstein}

While it is useful that LensPerfect can produce such detailed massmaps
when given so many multiple images,
what happens when LensPerfect is given far fewer constraints,
say a single quadruply-imaged galaxy?
Does the sum of four of the oddly-shaped basis functions depicted in
Fig.~\ref{KBF} yield a reasonable massmap solution?

\begin{figure*}
\plottwo{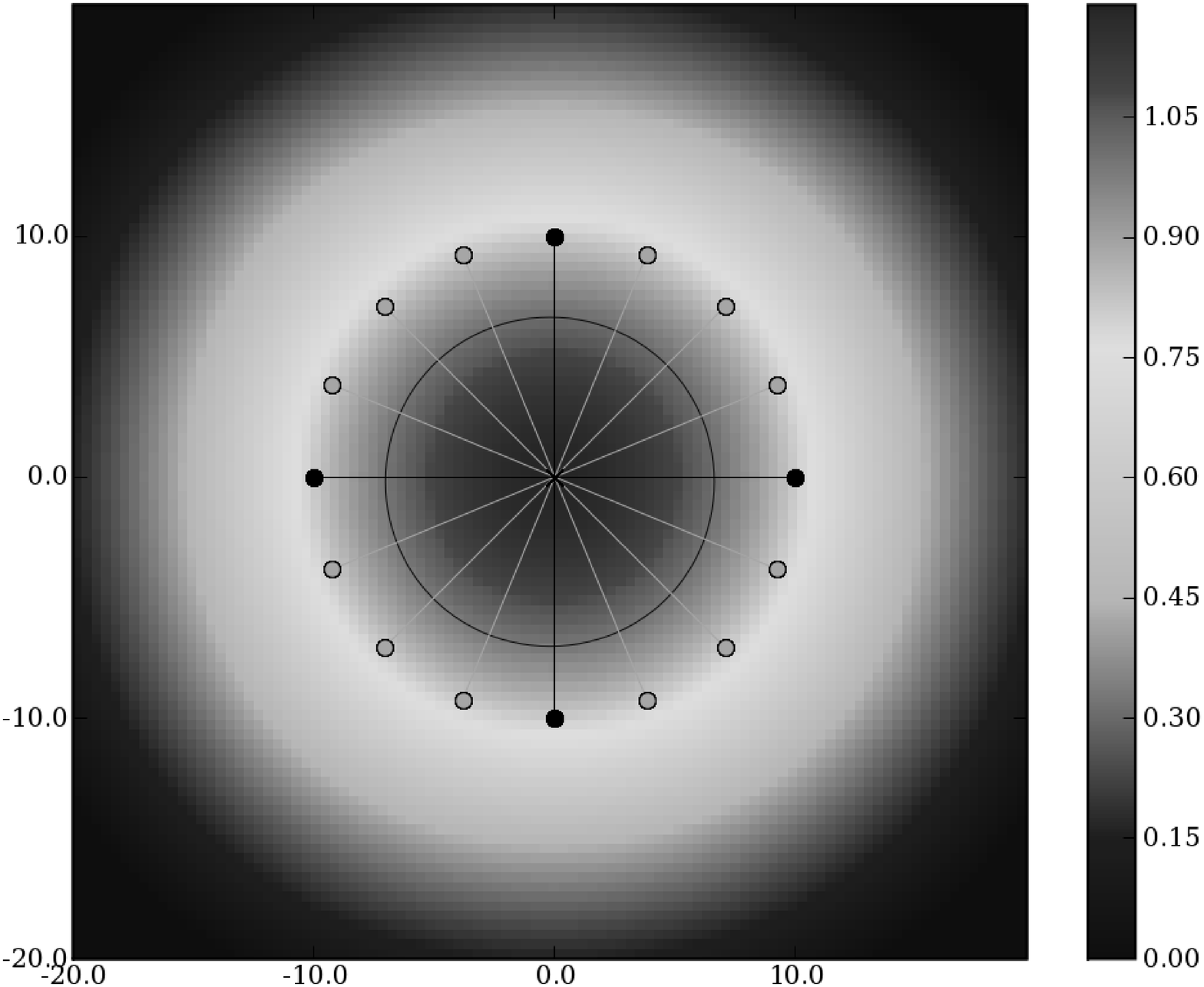}{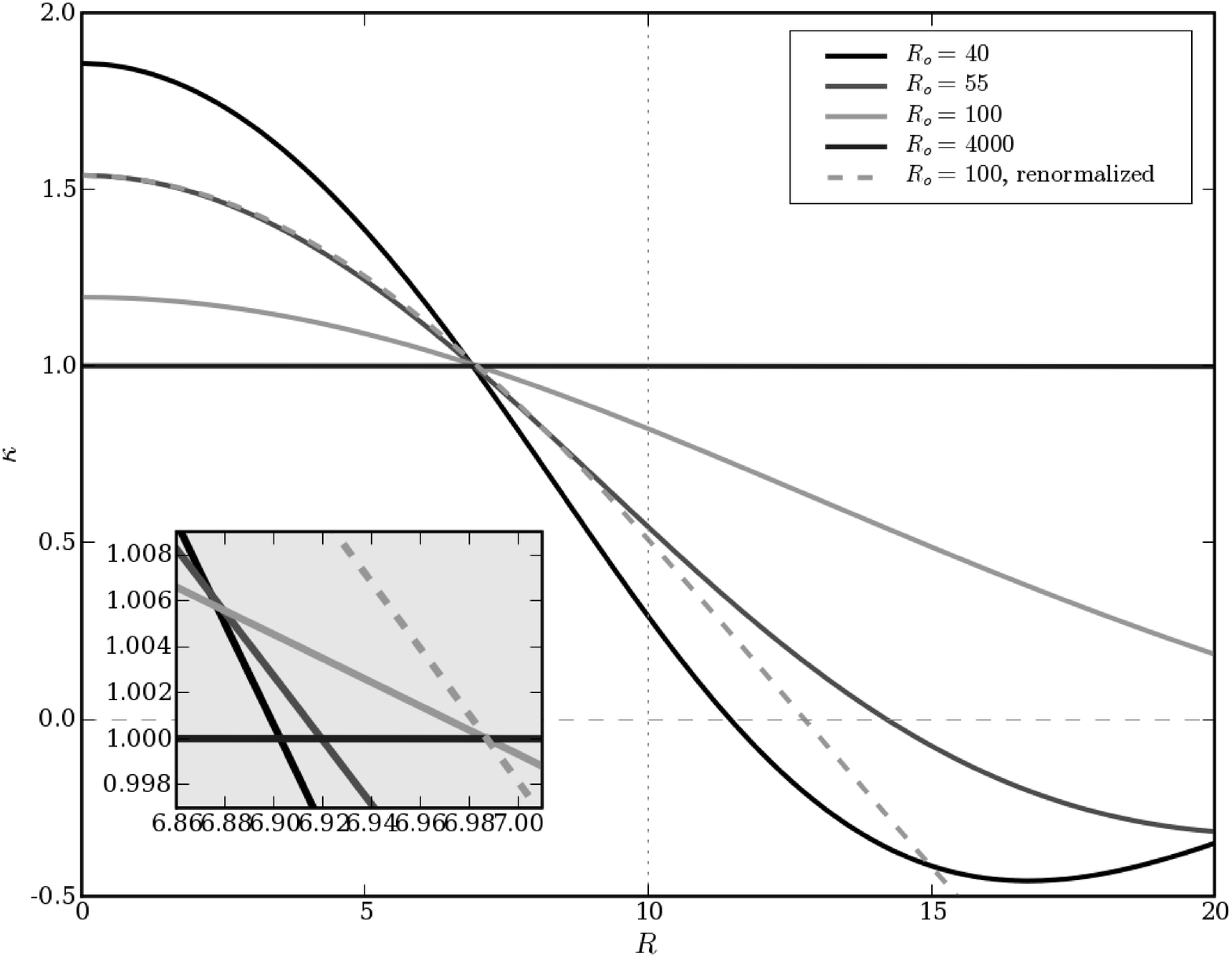}
\caption[Einstein ring solution and the mass-sheet degeneracy]{
\label{cross}
Left: LensPerfect massmap solution with $R_o = 100$ given an Einstein cross:
4 multiple images located at the black points,
and the source position at the center.
Given a more complete Einstein Ring (images at all 16 black and grey points),
the massmap solution is identical to within $\kappa = 0.01$.
The circle plotted inside the Einstein radius is the $\kappa = 1$ contour.
By changing $R_o$ (the width of our basis function),
we obtain solutions with different radial profiles,
the range of which is demonstrated by four examples plotted at right.
This demonstrates the classic ``steepness'' or ``mass-sheet'' degeneracy
$1 - \kappa^\prime = \lambda (1 - \kappa)$
almost exactly.
The dashed green line is the $R_o = 100$ solution
rescaled via this transformation
to match the peak value of the $R_o = 55$ solution.
The two line up nearly perfectly,
indicating that the degeneracies shown here are very similar to but not exactly
the classic form.
Also note in the zoom inset, that the different solutions
have $\kappa = 1$ at slightly different radii,
another indication that the classic mass-sheet degeneracy relation
is being followed approximately, but not exactly.
}
\end{figure*}

Yes it does.
Fig.~\ref{cross} (left-hand panel)
shows the massmap solution
obtained given a simple symmetric ``Einstein cross'' 4-image configuration.
The solution is exactly the same (to within $\kappa = 0.01$)
if we form a more complete ``Einstein ring'' with 16 image constraints 
as shown.
But this is just one possible solution, 
obtained with $R_o$ equal to ten times the Einstein Radius.
We begin to explore other solutions 
by varying the width $R_o$ of our basis function (Eq.~\ref{WendlandC6}).
Radial profiles of these different solutions are shown in
the right-hand panel.
By changing this single parameter, we neatly demonstrate
the ``steepness'' or ``mass-sheet''\footnote{As noted by 
\cite{SahaWilliams06},
the term ``mass-sheet'' degeneracy may lead to the mistaken belief 
that a constant mass sheet is what may be added without affecting the images.  
In fact the mass must also be rescaled, or multiplied, in concert.}
degeneracy
\citep[originally dubbed 
the ``magnification transformation'' in][]{Gorenstein88},
which states that the mass everywhere may be replaced by
\\
\begin{eqnarray}
\label{masssheet}
1 - \kappa^\prime & = & \lambda (1 - \kappa),\\
    \kappa^\prime & = & \lambda \kappa + (1 - \lambda)
\end{eqnarray}
\\
without affecting the observed image positions
(unless multiple galaxies of different redshifts have been lensed).
Note that the new mass $\kappa^\prime$ is steeper than the previous mass
$\kappa$ by a factor of $\lambda$.

The different LensPerfect solutions 
follow the transformation in Eq.~\ref{masssheet} 
very closely, although not exactly.
We have rescaled the green $R_o = 100$ curve via this transformation
such that its peak aligns with the $R_o = 55$ profile.
It now aligns with the full $R_o = 55$ curve very well
within the Einstein radius, but not perfectly.
(Note that the Einstein radius is also the convex hull in this case.
Thus we do not expect to be able to constrain the solution outside.)
Also note in the zoomed subplot, that the different profiles
attain $\kappa = 1$ at nearly the same radius $R$, 
but not exactly, as they would if they had resulted from
the transformation in Eq.~\ref{masssheet}.
Thus, the degeneracy we probe by simply varying $R_o$
is very similar to but slightly different than
the classic and simplest form of the ``mass-sheet degeneracy''.

We note that these solutions all have flat $\kappa(R)$ slopes at the center,
corresponding to a flat slope in $\rho(r)$, the three-dimensional density,
as well.
This slope gradually increases outward from the center.
This is the same behavior seen in many CDM simulations
\citep[and references therein,
although see \citealt{Diemand05}]{Navarro04, Merritt05, Merritt06},
which suggest that Dark Matter halos may attain 
the same \cite{Sersic68} density profiles 
that we observe in the light profiles of elliptical galaxies.
Thus the LensPerfect mass profiles obtained
when given only four image constraints
do appear to be reasonable.
But we stress that these solutions are far from unique,
and many other mass profiles are possible.

\subsection{Source Plane Shifts}
\label{massstick}

Theory tells us that given a single source
or even multiple sources at the same redshift,
the entire source plane can be shifted,
and a new massmap solution can be found
which does not affect the image positions.
The degeneracy was originally named the ``prismatic transformation'' 
by \cite{Gorenstein88}, who explained that
this shift can be accomplished by adding an infinitely long and thin mass stick
to the solution off to one side of the images.
Of course such mass sticks are not very physical.
Thus, we can overcome this degeneracy by simply finding that solution 
which does not resort to the addition of mass sticks!
This is why we ``know'' that
Einstein rings are produced by an on-axis alignment 
of the lens and source galaxies.
The on-axis configuration requires a simple symmetric massmap,
while any off-axis source position would require a less physical massmap,
perhaps requiring the addition of a mass stick.

\begin{figure}
\plotone{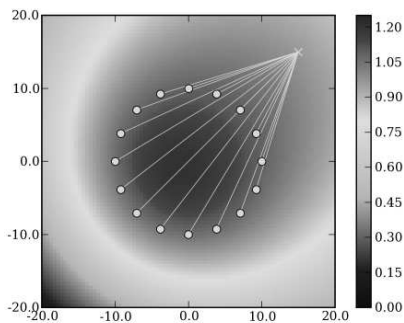}
\caption[Einstein ring with source position offset]{
\label{ringoffset}
Einstein Ring produced by a galaxy offset from the ring.
While this configuration is possible, 
it is highly unlikely that an offset source galaxy 
and asymmetric lensing mass distribution
would conspire so precisely to produce an Einstein ring in Nature.
But do note that the massmap solution inside the ring is nearly identical 
to that obtained with the source position placed at the center of the ring.
}
\end{figure}

We say ``perhaps'' because LensPerfect is able to find solutions
to such off-axis Einstein rings without resorting to mass sticks.
(Fortunately LensPerfect has no mass-stick basis functions to work with.)
One such solution is shown in Fig.~\ref{ringoffset}.
This mass distribution, while more plausible than a mass stick,
is nevertheless highly improbable.
The proper amount of mass must be added off-axis in the correct configuration
to refocus the light just enough toward the center of the lens
where it then produces the Einstein Ring.
It is very unlikely that such a precisely-tuned mass configuration
would occur in Nature.
Much more likely is the on-axis alignment,
in which case any simple symmetric mass configuration will do.
Our optimization procedure (\S\S\ \ref{sourcepos}, \ref{optimization})
quickly converges to the on-axis source position as being most likely,
as off-axis solutions are penalized for being asymmetric.

But we should not be surprised to find modest source position shifts 
in our solution either (\S\ \ref{sourcepossec}).
The mass within the convex hull is extremely insensitive to these shifts.
(A constant shift in the source plane 
does not affect the divergence of the deflection field 
at the multiple image positions.
Thus the massmap within the enclosed region 
experiences only negligible changes.)

\subsection{Constraining Image Shapes and Sizes}
\label{knots}

Additional constraints may be derived from gravitationally lensed images
by requiring the mass model to correctly reproduce not only image positions,
but also their sizes, shapes, orientations, and fluxes.
A method to constrain the fluxes (and shears) of unresolved images
was discussed in \S\ \ref{fluxes}.
But observed fluxes (and especially shears) may be uncertain.
More precise constraints may be obtained when the multiple images
are resolved and extra knots may be identified within them.

\begin{figure*}
\plottwo{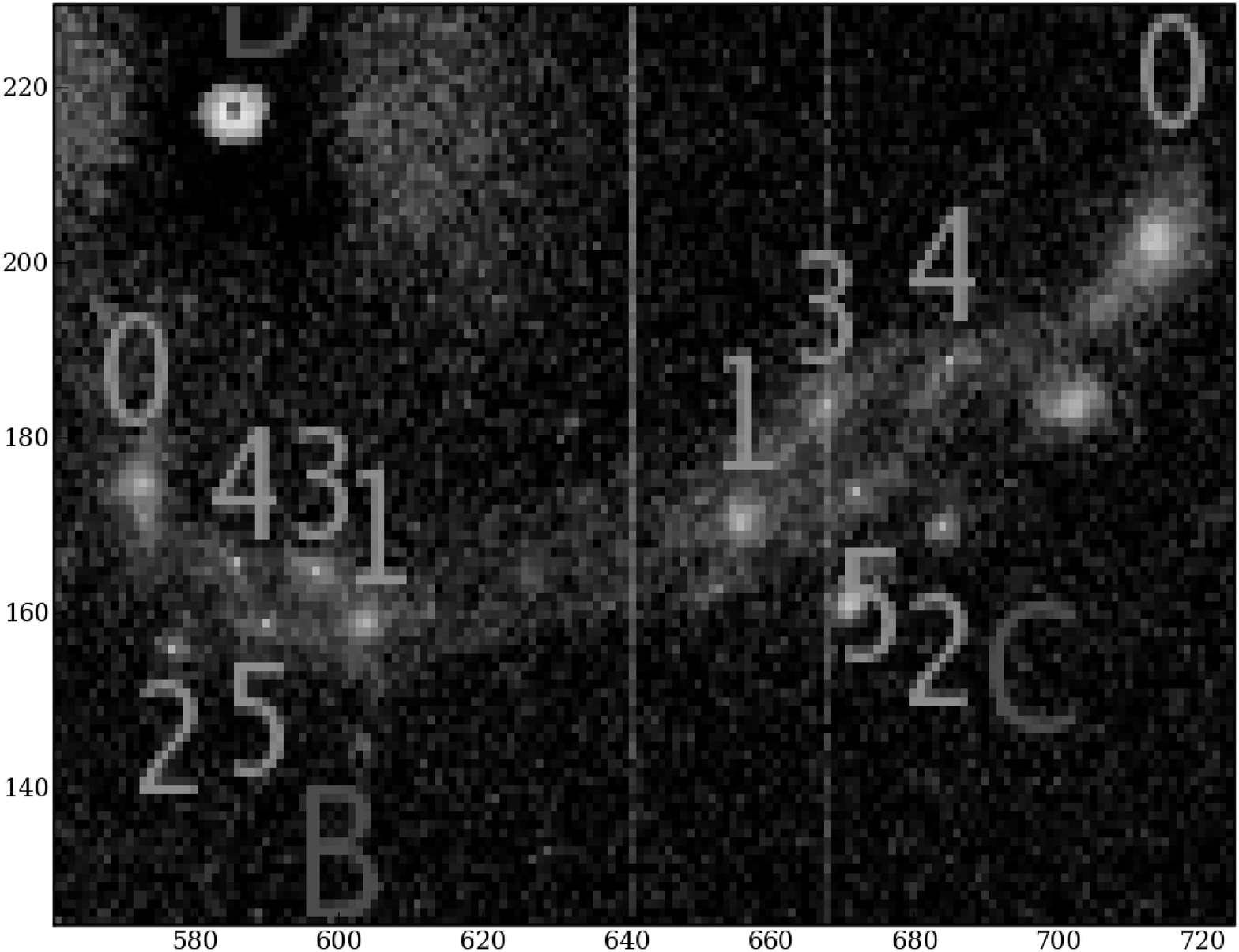}{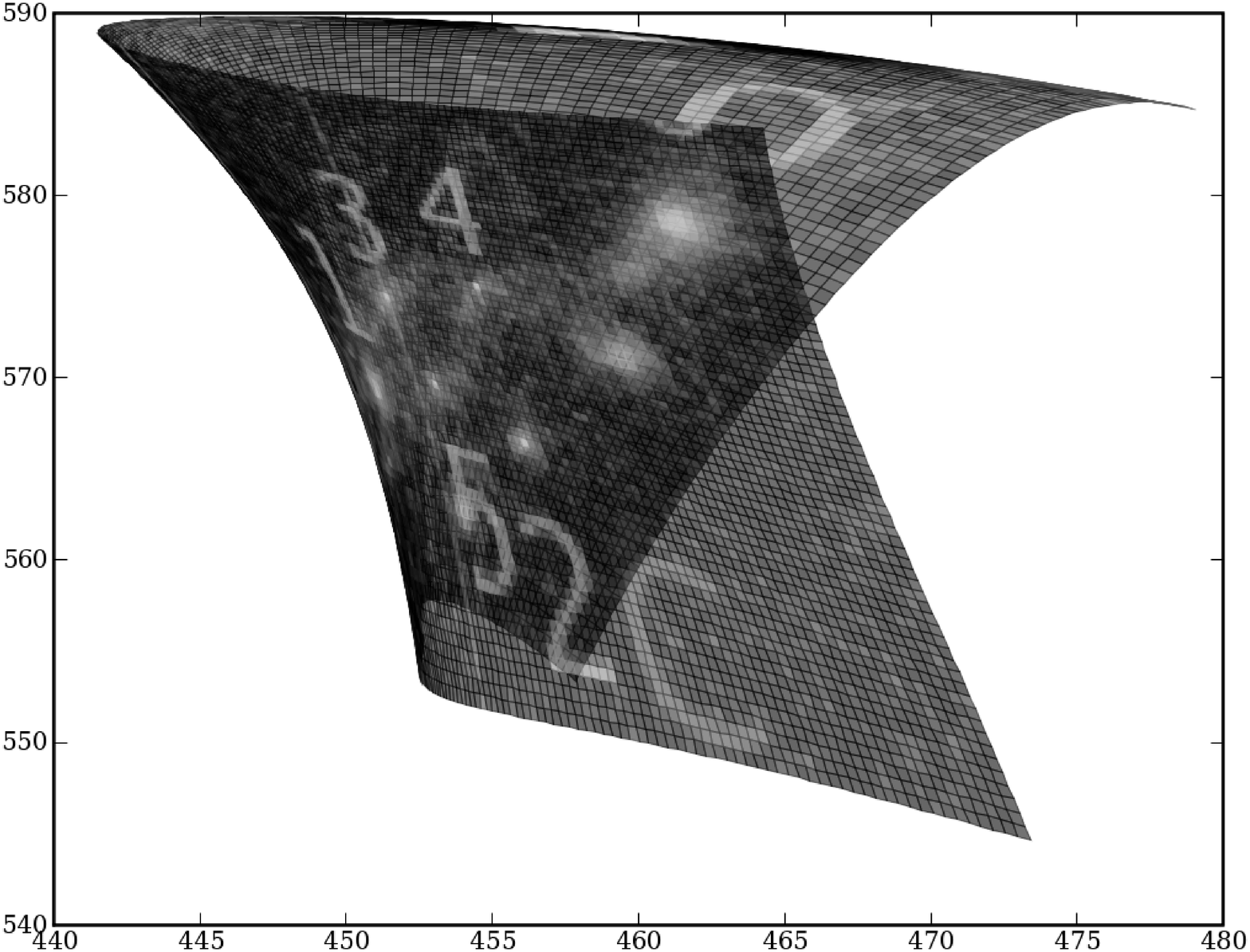}
\caption[MS1358 fold images B and C, delensed so that knots overlap]{
\label{MS1358BC}
Left: Two images of a quadruply-lensed $z=4.92$ galaxy behind MS1358.
Six matching knots are identified and labeled in each of these two images.
A cluster galaxy has been subtracted from this $BVi\arcmin z\arcmin JH$
color image.
Right: The same image de-lensed back to the source plane
given the LensPerfect massmap solution.
All knots have been constrained to align in the source image.
(Also see Fig.~\ref{MS1358BC2}.)
}
\end{figure*}

\begin{figure*}
\plottwo{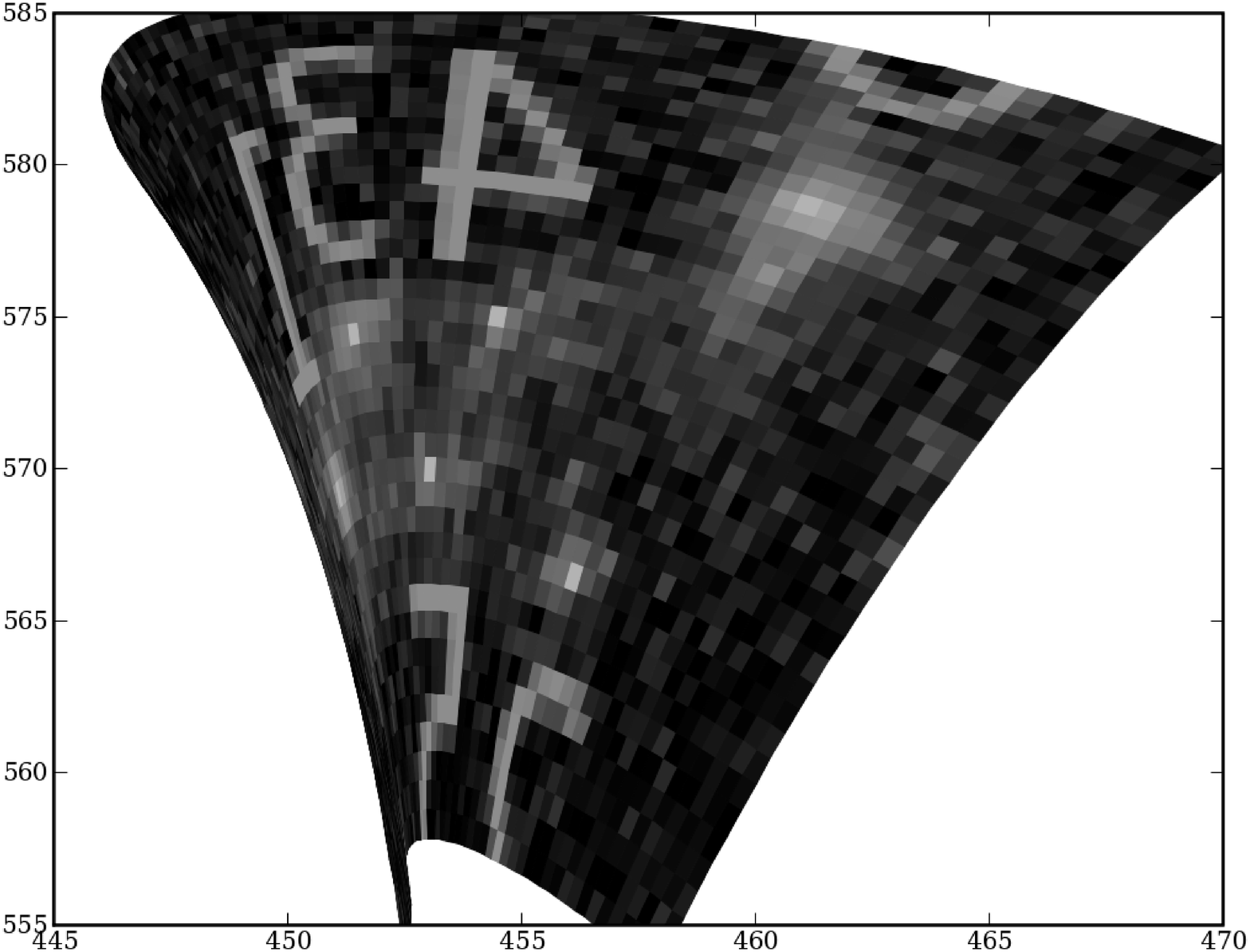}{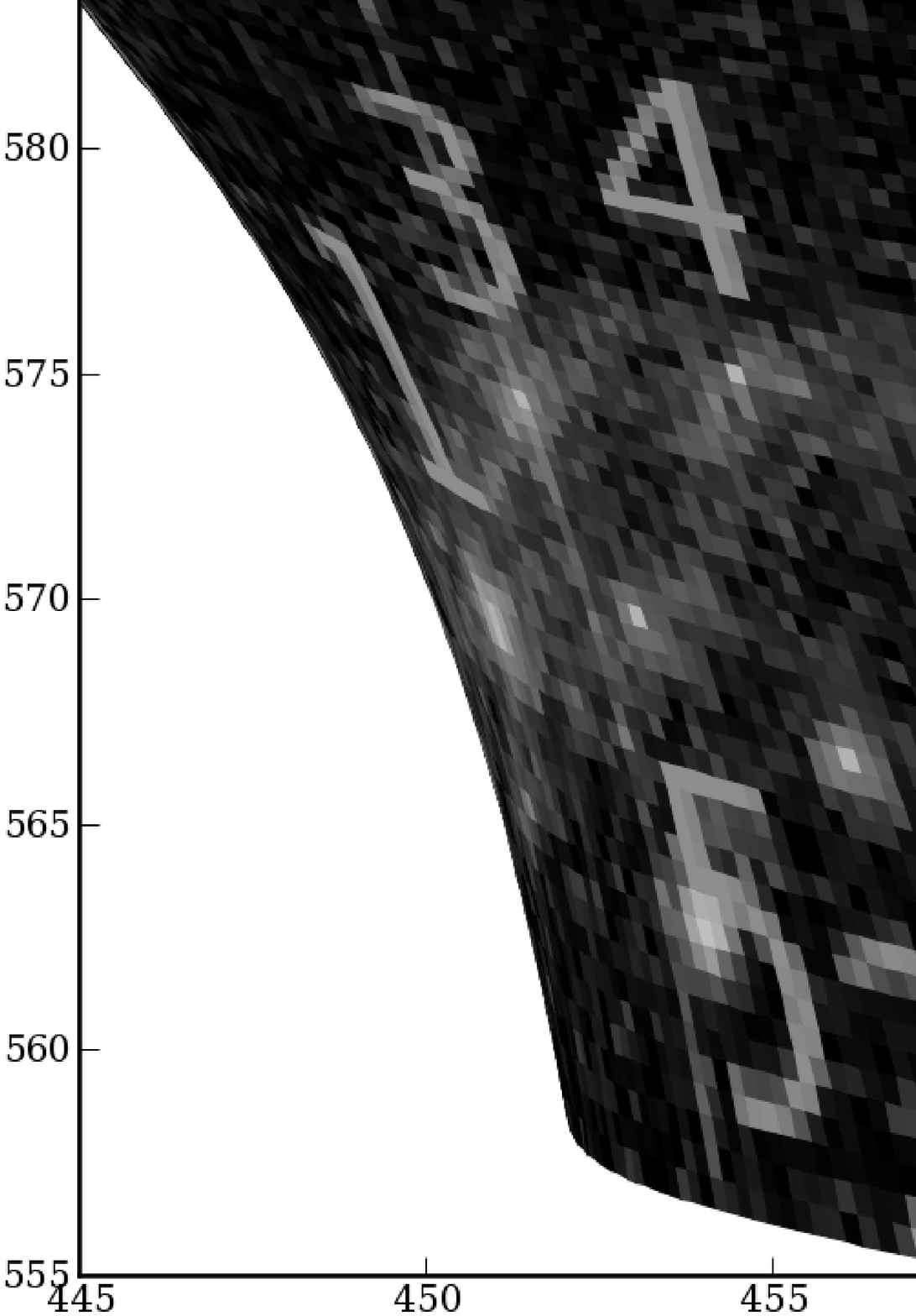}
\caption[MS1358 fold images B and C, delensed]{
\label{MS1358BC2}
Images B and C de-lensed back to the source plane using LensPerfect.
The six knots have been constrained to align exactly in the de-lensed images.
}
\end{figure*}

Fig.~\ref{MS1358BC} shows two of the four multiple images 
of a $z = 4.92$ galaxy 
produced by the galaxy cluster MS1358 \citep{Franx97}.
Six knots are identified as common in both of these images (B \& C),
with three of the knots also being identified in image A (not shown).
For each knot in turn, all of its images are constrained
to originate from the same source position.
The deflection model is updated after each knot is added.
The final deflection solution yields the source plane images in 
the right-hand panel (see also Fig.~\ref{MS1358BC2}).
Note the identical alignment of the two images (B \& C).
At this point, the de-lensed images may be co-added 
to obtain greater depth, if desired \citep[e.g.,][]{Colley96}.

\begin{figure*}
\plottwo{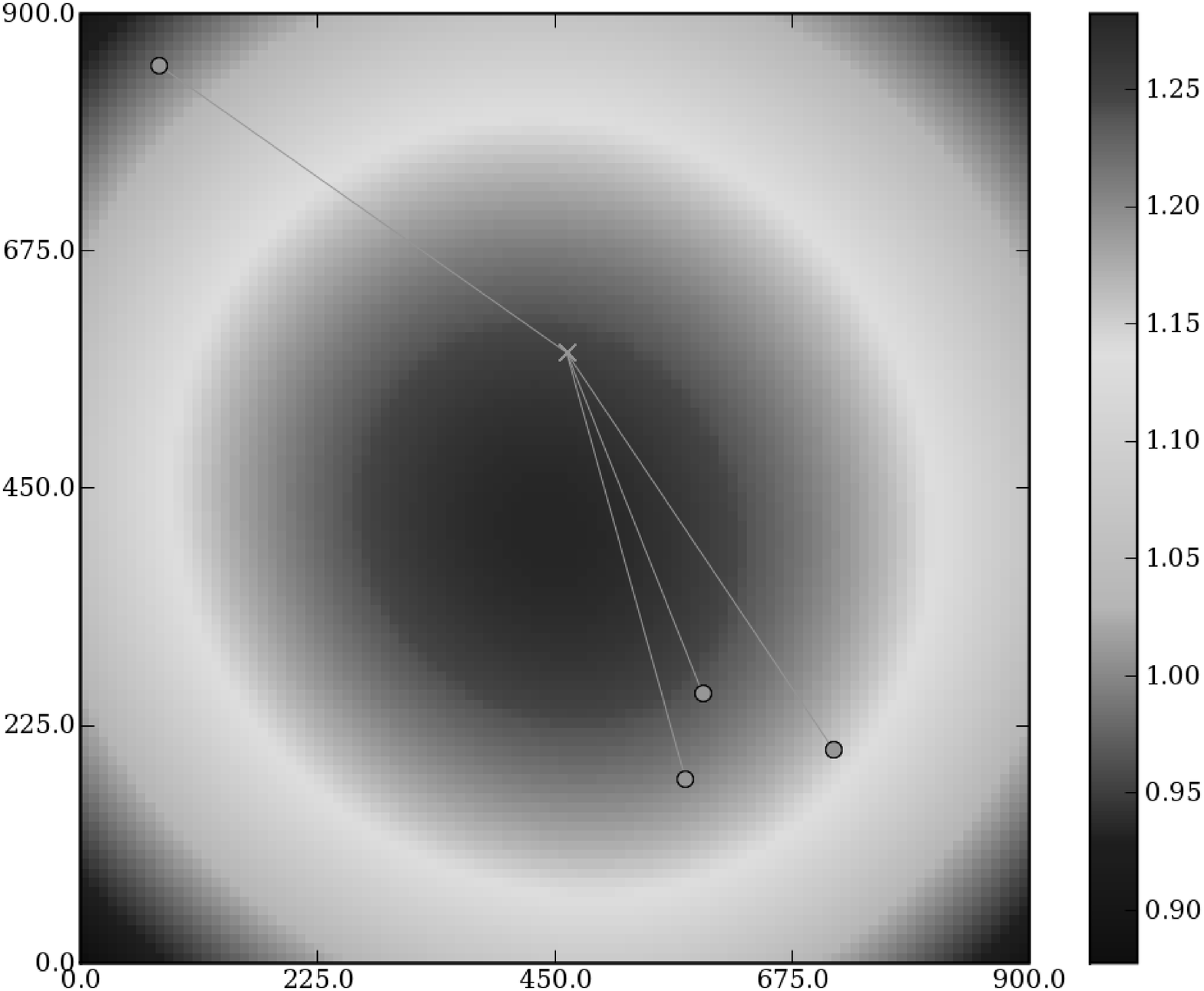}{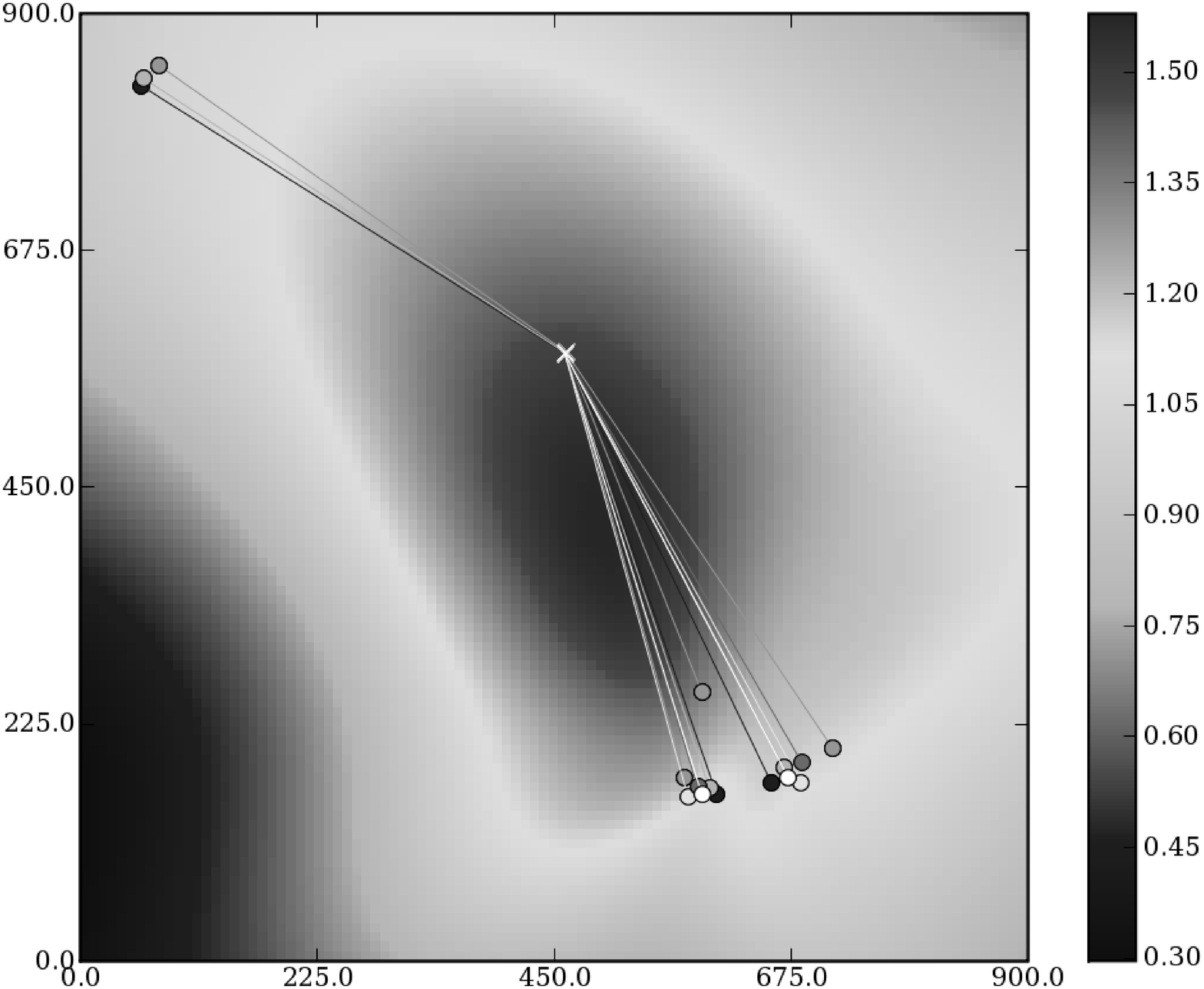}
\caption[MS1358 massmap improvement from adding knot constraints]{
\label{MS1358k}
MS1358 massmap solutions obtained by LensPerfect.
Left: Only the positions of the brightest knots are constrained.
Right: The positions of six matching knots are constrained in images B and C
along with three knots in image A and one in image D.
Note that the colormaps in the two plots do not have the same scale.
}
\end{figure*}

But this capability allows us to do more than simply
produce pretty de-lensed images.
The extra constraints provided by image knots 
can add significant detail to the massmap solution.
Fig.~\ref{MS1358k} compares 
(left) the MS1358 massmap solution obtained
when using only a single image position (the brightest knot) for each object
and (right) the solution obtained when constraining all knot positions.
In the latter, the mass is more stretched toward a second cluster galaxy
which proves crucial to traditional parametric mass modeling.

\section{Discussion}
\label{discussion}

We now describe the current limitations of the LensPerfect method
and plans for future capabilities LensPerfect may provide us with.
We also discuss the relative merits of both 
``parametric'' and ``non-parametric'' methods,
elaborating on points made in the introduction.

\subsection{Imperfect Perfect Solutions}

The ``perfect'' in LensPerfect refers to the 
reproduction of multiple image positions attained by its massmap solutions.
But these solutions may be perfectly wrong and even unphysical
if incorrect source positions and/or redshifts 
(and/or multiple image identifications) are provided.
We have developed a method to find source positions and redshifts
which produce reasonable solutions.
We do not claim this method is perfect;
it appears to work well but may be improved upon in the future.

LensPerfect massmap solutions do not guarantee against 
predictions of additional multiple images which don't exist.
This is an issue common to many methods,
and it has been argued alternately
that this should either more or less be a concern.
The argument against its importance
is that the prediction of additional multiple images 
is very sensitive to local substructure.
Slight modifications might be made to this substructure,
it is argued, to make these extra predicted multiple images disappear.
LensPerfect may finally give us a tool capable of easily demonstrating this.
We may be able to add deflection field constraints 
that tweak the massmap just enough to eliminate unwanted images.
This idea has not yet been tested.
In the meantime, we will simply check each solution
for the correct number of multiple images.
We find our models generally do not produce extra multiple images,
as long as reasonable source positions are assumed.

\subsection{Imperfect Solutions Superior?}
\label{imperfect}

Would it help to loosen the restriction that our massmap solutions
perfectly reproduce all the multiple image positions?
That is, might we obtain more accurate solutions
by allowing our solutions to be imperfect?
There are two parts to this question.

First we consider measurement uncertainties.
We have already discussed how we deal with redshift uncertainties in 
\S\ \ref{optimization}.
We allow the redshift to vary within the uncertainties
as we optimize the massmap.
The same could be done for other uncertain measurements.
Image position uncertainties are generally very small (about one pixel).
But when adding flux / shear constraints,
we should certainly include the corresponding uncertainties in our optimization.
\cite{EvansWitt03} and \cite{CongdonKeeton05} followed this approach
in their studies of galaxy lenses.
By including uncertainties, they noted modest improvement in their solutions.

The second part of the question is more fundamental.
Could a less perfect massmap be more accurate?
This does prove true in the SLAP method \citep{Diego05a}.
They purposely leave some scatter in the source plane to avoid
solutions which are ``biased'' with ``a lot of substructure''.
But this is likely a result of their formulation of the problem.
They consider all pixels in each lensed image
and find that massmap solution 
which minimizes the {\em sizes} of all the delensed images.
Of course the delensed sizes should not be zero,
so the delensed image pixels are allowed some ``scatter''.
(They make no attempt to match internal features 
as we demonstrate in \S\ \ref{knots}.)
Meanwhile, the proponents of PixeLens 
do not report such problems with their solutions
which perfectly reproduce all image positions \citep{Saha06}.

We performed a test to see what an imperfect solution would look like
and if it might be more accurate.
We began with our perfect solution 
to the 93 multiple images in \S\ \ref{A1689model}
with optimized source positions.
We then ``reoptimized'' the source positions 
(based on our physicality criteria), 
one-by-one {\em for each individual image}.
Source positions were no longer constrained 
to a single point for each multiple image system.
In practice, they drifted by an average of 0.4 pixels.
Our resulting imperfect massmap was not extremely different
from our perfect massmap
but not any more accurate either.
Some of the substructure features in the center were erased
as the physicality measure guided the optimization toward a smoother solution.
The radial profile did not significantly improve nor deteriorate in accuracy.

We believe the ``1,000 points of light'' massmap in Fig.~\ref{kappa1000}
is a convincing demonstration that perfect is best.
When the deflection field is sampled at 1,000 points,
we are able to map fine detail in the massmap.
This fine structure is encoded in the exact positions of the multiple images.
Allowing for an imperfect solution would erase this fine structure
and waste information obtained in the observations.

\subsection{Weak Lensing, Extended Images, and Time Delays}
\label{weaklensing}

It may be possible to directly incorporate 
measurements of both ``weak'' lensing and not-so-weak shear
into the LensPerfect method.
Lacking multiple identifiable image knots,
we can constrain galaxy shapes
using the same method we use to constrain fluxes
(\S\ \ref{fluxes}).
Instead of altering the relative sizes of the source and lensed regions,
we can alter the shear.
Thus we could constrain all delensed images to be round,
or perhaps round with some scatter of ellipticity 
as measured in large scale surveys.
We do not develop this idea further here.
Such a technique would require three constraints per galaxy,
which, given 100 galaxies, would require significant computing time
(although at this stage iterations may not be required).

A more practical idea may be to compare measured shears
to those predicted from our mass models
and include the disagreement as a penalty in our optimization procedure.
But note that this would only work well for galaxies within our convex hull
as our solution (and thus shears) would be very poorly constrained 
outside this region.

Beyond simple constraints of flux, shear, and image knots,
all of the pixel information in each lensed image
may be utilized in the model derivation 
\citep[e.g.,][]{WarrenDye03, Suyu06, Koopmans06}.
This is especially desirable in strong galaxy-quasar lensing,
in which observable constraints are less plentiful than in cluster lensing.
It is unclear how to implement this in LensPerfect,
except as another penalty in our optimization routine.

Another constraint often used in strong galaxy-quasar lensing studies
is time delays observed among different images.
Time delays are not local functions of the deflection field,
and as such, we cannot constrain them directly with LensPerfect
as we constrain fluxes and shears.
But for a given solution,
they can be calculated directly from derivatives of our basis functions
and compared to those observed,
yielding another penalty for our optimization routine.

\subsection{``Parametric'' vs.~``Non-parametric'' Methods}

Methods are often classified as ``parametric'' or ``non-parametric'',
depending on whether a clear physical parameterization is used to 
construct the proposed massmap solutions.
``Grid-based'' methods are often referred to as ``non-parametric'',
even though strictly speaking they do have parameters, 
namely the mass at every pixel on a grid.
The real distinction to be made here is between
``model-based'' and ``model-free'' methods.
The former construct mass halos as physical analytical forms,
while the latter do not.
To give examples,
the fully model-based strong lensing analyses of the A1689 ACS images
published to date have been \cite{Halkola06, Halkola07} and \cite{Limousin07}.
Meanwhile, \cite{Diego05}, \cite{Saha06}, and \cite{Leonard07} 
have obtained model-free massmaps based on these images.
The \cite{Broadhurst05} and \cite{Zekser06} analyses
included both model-based and model-free elements.

LensPerfect, while clearly parametric, is also model-free.
The LensPerfect solutions are given as sums of basis functions.
But these basis functions have no physical interpretation.
And this functional form is practically indiscernible in the final solutions.
In fact, the basis function coefficients are generally
many orders of magnitude greater than the amplitude of the massmap.
These large mass components all cancel out nearly perfectly
with the ``residuals'' being the massmap solution.

But while LensPerfect is model-free, 
it does not have the large number of free parameters 
typical of grid-based methods.
In fact when fitting image positions only
(including extra knots, but not fluxes),
the number of free parameters solved for (the coefficients)
is exactly equal to the number of constraints.
Note that the source positions and redshifts are not solved for directly,
but rather must be provided as input.
Each flux measurement provides a single constraint
but two free parameters are required to constrain it in our models.
And a shear measurement provides two constraints,
which can be reproduced with an equal number of (two) free parameters.

But which are superior overall, model-based or model-free methods?
Individual researchers may have a decided preference for one over the other,
but in fact 
both types of methods have their strengths and each serves a purpose.

Physical model-based massmaps allow us to test whether Dark Matter
is distributed in certain ways, according to our assumptions.
In principle, they allow for a more straightforward determination
of meaningful physical parameters.
Cluster mass models attempt to simultaneously constrain the forms of both 
the overall cluster halo and the individual galaxy halos
\citep[e.g.,][]{Halkola07}.
But degeneracies are strong between the two additive components.

On the other hand, model-free massmap reconstruction methods 
allow us to directly test for the presence of Dark Matter
free of assumptions about its distribution.
In particular, these methods make no assumptions about mass following light.
Some of our most important discoveries about Dark Matter
have come and are expected to come from cases in which
mass does {\it not} follow light.
Mass peaks offset from light peaks can provide constraints
on the collisional nature of Dark Matter particles,
as in the Bullet Cluster \citep{Markevitch04, Clowe06}.
And observations of Dark substructure (not associated with light)
may someday vindicate CDM simulations 
which predict much more halo substructure than is visible 
\citep[and references therein]{Diemand07, Strigari07}.

Model-free methods are also able to explore a wider range of 
possible massmap solutions, including (with the advent of LensPerfect) 
those which perfectly reproduce all 100+ multiple image positions.
Of course, exploring the full range of model-free solutions
can be a computational challenge.
And it is not entirely clear how to sort the physical solutions 
from those which are ``less'' physical or aphysical.

Meanwhile, too much model flexibility has been cited a potential problem,
as a flexible model may fit incorrect data without any alarms sounding.
\cite{Limousin07} claim that their model-based method
and that of \cite{Halkola06} were unable to fit some multiple image systems
incorrectly identified in the initial \cite{Broadhurst05} work,
which was a bit more model-free.
But LensPerfect, while more flexible still,
may actually be more unforgiving than all previous methods
when it comes to incorrect multiple image identifications.
The reason is simple.
``Imperfect'' massmap reconstructions
experience some offset in each and every predicted image position.
Thus a misidentified multiple image set may have a larger $\chi^2$ than average,
but this may be more easily dismissed in the analysis.
In LensPerfect, however,
each multiple image puts a rigid constraint on the deflection field.
Thus a misidentified multiple image is more likely to
cause the deflection field to get tangled,
leading to a ``less physical'' (if not aphysical) solution.
We stay alert for such ill-fitting multiple image systems
as we add each to our models.
And, of course, we must take care to avoid misidentifying multiple images
from the start whenever possible.

Another objection to model flexibility was raised by \cite{Kochanek04_review}. 
He takes exception to the ability of
the model-based but flexible \cite{EvansWitt03} method
to produce a mass model that accounts for the flux anomaly observed 
in the strong galaxy-quasar lens Q2237+0305
even though this flux anomaly has since been shown 
to have been due to a microlensing event, which has now passed!
We would argue that the mass model proposed by \cite{EvansWitt03}
is but one possible solution among several to the observed flux anomaly.
All possible solutions should be considered,
and in this case, microlensing is proven to be the true solution.

In analyses of galaxy-quasar lensing,
model-based methods enjoy even greater appeal
than in cluster lensing.
Perfect solutions are much easier to come by 
when there are fewer constraints (4 image positions, for example).
And with so few image constraints,
a large range of solutions is possible.
We demonstrated one way of probing this solution space in \S\ \ref{Einstein},
and the PixeLens method provides another.
But many choose to slash the solution space
by making well-motivated model-based assumptions (e.g., an isothermal profile).

A powerful technique would be to combine LensPerfect with a model-based method.
We could find good (imperfect) model-based solutions 
which leave offsets between the observed and predicted image positions,
and then perfect these solutions with LensPerfect.
The deflection offsets may be interpolated, and this ``offset solution''
added to the imperfect solution to create a perfect solution.
We implemented this idea, but the results appeared unruly.
A more creative approach will be required
if this capability is to be realized.

We mention one other advantage to our method.
That advantage is ease of use.
If a simple mass model is required, LensPerfect's speed is hard to beat.
Given a single lensed galaxy, 
LensPerfect instantly obtains a perfect solution ``out of the box''.
And if instead given 30+ lensed galaxies as in Abell 1689,
LensPerfect still obtains solutions with minimal user input.
Parametric methods instead require the user to develop complex models
capable of fitting so many multiple images well.
Previous studies identified and measured properties of
many cluster galaxies
in order to model a ``galaxy component'' 
which would be added to a separate halo component
in their solutions.
LensPerfect instead bypasses this parameterization process
obtaining a detailed massmap free of strong assumptions.

\section{Summary and Future Work}
\label{summary}

We have presented a new approach to gravitational lens massmap reconstruction.
Given image positions, source positions, and redshifts,
a new mathematical technique is used to
interpolate the deflection field via direct inversion.
The resulting massmap (simply half the divergence of the deflection field)
perfectly reproduces all of the observed image positions.

In practice, source positions are unknown.
We have devised a method that efficiently optimizes 
over different possible configurations of the source galaxies.
Each configuration produces a different massmap solution
which is evaluated for ``physicality'' 
based on criteria developed as part of this work.
Our criteria make only minimal assumptions about the form of the massmap.
Specifically, they make no assumptions about the slope of the radial profile
nor mass following light.

We demonstrated our method on mock gravitational lensing data.
A known massmap was used to lens 19 mock galaxies to produce 93 multiple images.
Our massmap solution perfectly reproduces all the multiple image positions
while accurately recovering the known lens mass distribution
to a resolution limited by the density of the multiple images.
We demonstrate the improved accuracy and fine detail we may expect
from a massmap derived from 1,000 multiple images.

We also presented LensPerfect solutions based on far fewer constraints,
such as four multiple images of a single galaxy.
Image fluxes or extra knots may provide additional constraints
which can also be perfectly fit by our models.
The number of free parameters solved for 
is kept nearly equal to the number of observable constraints.

The LensPerfect software is easy to use
and is made publicly available at our website (see \S\ \ref{intro}).

In subsequent papers, 
we will apply this method to Abell 1689 and other galaxy clusters.
We will also attempt to resolve some of the ``flux anomalies'' 
observed in galaxy lenses.

In \S\ \ref{discussion}, 
we discussed some possible improvements and extensions of our method,
including the incorporation of weak lensing data.
We also hope to better quantify the accuracy of our massmap recovery
and compare our performance to that of other methods.
The uncertanities in our solutions should be well determined
by exploring the full range of physical solutions.
We will also experiment with other radial basis functions
and other recipes for measuring massmap physicality.

\acknowledgements
We would like to thank many 
for useful discussions during the development of LensPerfect including 
Justin Read, 
Rick White, Marijn Franx, Piero Rosati, Myungkook Jee,
Aleksi Halkola, Stella Seitz, Ralf Bender,
Jean-Paul Kneib, Bernard Fort, Genevieve Soucail,
Tony Tyson, Chris Fassnacht, Leonidas Moustakas, and Maru\v{s}a Brada\v{c}.
We thank Arjen van der Wel for comments that helped improve the manuscript.
And we especially thank our anonymous referee for 
a thorough reading of the manuscript and useful scientific contributions.
This research is supported by the 
European Commission Marie Curie International Reintegration Grant 017288-BPZ
and the PNAYA grant AYA2005-09413-C02.

\appendix
\section{Massmap Physicality Penalty Function}
\label{penalty}

To calculate our penalty function, 
we first evaluate the massmap solution on a $41 \times 41$ grid. 
This proves large enough to measure physicality, 
while taking less than a second to calculate 
given as many as 50 image positions.
(As this evaluation is to be part of our 
iterative procedure to determine source positions, 
it should be kept as quick as possible.)
Based on this $41 \times 41$ massmap, we calculate
the total penalty as follows:

\begin{enumerate}

\item The sum of all negative pixels inside the convex hull 
is multiplied by $-100$.
(We could simply assign a penalty of infinity to negative massmaps, but this 
would create a large region of constant penalty in source coordinate space. 
Our finite and varying penalty serves better during the optimization 
to corral the source positions 
back toward those which yield positive solutions.)

\item The mean mass $\langle \kappa \rangle$ 
and RMS scatter $\Delta \kappa$
are measured in radial bins of 80 points each. 
The RMS scatter is totaled and divided by 10. 
Within the convex hull, the mean and scatter are recalculated 
in radial bins of 40 points each. 
This inner scatter is totaled and multiplied by 4.
Outside the convex hull, we rebin the massmap, yet again,
in bins of 80 points each. 
Rather than the RMS mass scatter, 
this time we penalize the peak-to-peak variation
(that is, the maximum minus the minimum) in each bin. 
These variations are totaled and multiplied by 5 
and divided by the number of bins.

\item Any outward increase in $\langle \kappa \rangle$ over $R$ 
is measured and multiplied by 10.
This is calculated both for all points 
and then again for those within the convex hull.
Both penalties are added.

\item Given the mean mass $\langle \kappa \rangle$ 
and RMS scatter $\Delta \kappa$
calculated within the convex hull (see 2 above), 
we interpolate the 1-$\sigma$ lower limit 
$\langle \kappa \rangle$ - $\Delta \kappa$
for all points within the convex hull.
Deviation below this limit is measured and divided by 2.
Note that we certainly expect some points to fall below 
the 1-$\sigma$ lower bound. 
(By definition, 16\% of the points should fall below.) 
But the idea is to minimize these deviations.
A large deviation below indicates a ``tunnel'' or unphysical dip 
in the surface density.

\item Finally, we put a premium on smooth solutions within the convex hull 
as being the most likely. 
Numerical derivatives of the massmap are calculated at each pixel
($x$, $y$):
$d\kappa / dx = \vert \kappa(x+1, y) - \kappa(x-1, y) \vert$,
$d\kappa / dy = \vert \kappa(x, y+1) - \kappa(x, y-1) \vert$
The absolute values of these are totaled the entire sum and divided by 5.

\end{enumerate}

The above penalties and their respective weights were 
defined after much trial and error. 
We find that massmaps with lowest total penalty defined as above appear
to be the best behaved, or most physical. 
Again, other weights and penalty schemes are certainly possible.

\bibliography{}

\end{document}